\documentclass[acmsmall,screen,nonacm=true,review=false,natbib=false]{acmart}

\setcopyright{acmlicensed}
\copyrightyear{2025}
\acmYear{2025}
\acmDOI{XXXXXXX.XXXXXXX}

\acmJournal{TOMS}
\acmVolume{0}
\acmNumber{0}
\acmArticle{0}
\acmMonth{0}

\usepackage[export]{adjustbox}

\usepackage{listings}
\usepackage{array}
\usepackage[inline]{enumitem}
\usepackage{subcaption}
\usepackage{nameref}
\usepackage{siunitx}

\usepackage[capitalise, noabbrev]{cleveref} %

\usepackage{tikz}
\usetikzlibrary{matrix,arrows.meta, calc, positioning, decorations.pathreplacing, patterns, shapes}

\RequirePackage[backend=biber,
                datamodel=acmdatamodel,
                style=acmnumeric
               ]{biblatex}
\newcommand{\code}[1]{\texttt{#1}}
\newcommand{\mirge}{\textsc{MIRGE}}
\newcommand{\mirgecom}{\textit{MIRGE-Com}}
\newcommand{\software}[1]{\textit{#1}}
\def\CPP{{C\nolinebreak[4]\hspace{-.05em}\raisebox{.4ex}{\tiny\bf ++}}}

\makeatletter
\newcolumntype{M}[1]{>{\centering\arraybackslash}m{#1}}
\newcolumntype{L}[1]{>{\raggedright\arraybackslash}m{#1}}
\makeatother

\definecolor{codegray}{gray}{0.95}

\definecolor{illiniorange}{RGB}{255,95,5}
\definecolor{illiniblue}{RGB}{19,41,75}
\definecolor{altgeld}{RGB}{200,65,19}
\definecolor{industrial}{RGB}{29,88,167}
\definecolor{prairie}{RGB}{0,98,48}
\definecolor{storm1}{HTML}{707372}
\definecolor{storm2}{HTML}{9C9A9D}
\definecolor{storm3}{HTML}{C8C6C7}

\definecolor{myblue}{named}{industrial}
\definecolor{myred}{named}{altgeld}
\definecolor{mygreen}{named}{prairie}
\definecolor{mygray}{named}{storm1}

\definecolor{tikzred}{named}{altgeld}
\definecolor{tikzblue}{named}{industrial}
\definecolor{tikzgreen}{named}{prairie}
\definecolor{tikzgray}{named}{storm1}
\definecolor{tikzlightgray}{named}{storm3}

\begin{document}

\title{\mirge: An Array-Based Computational Framework for Scientific Computing}

\author{Matthias Diener}\email{mdiener@illinois.edu}
\orcid{0000-0002-9064-7806}
\author{Matthew J.~Smith}\email{mjsmith6@illinois.edu}
\author{Michael T.~Campbell}\email{mtcampbe@illinois.edu}
\author{Kaushik Kulkarni}
\orcid{0009-0001-0645-2169}
\email{kgk2@illinois.edu}
\author{Michael J. Anderson}\email{manders2@illinois.edu}
\author{Andreas Kl\"ockner}\email{andreask@illinois.edu}
\orcid{0000-0003-1228-519X}
\email{andreask@illinois.edu}
\author{William Gropp}\email{wgropp@illinois.edu}
\author{Jonathan B.~Freund}\email{jbfreund@illinois.edu}
\author{Luke N.~Olson}\authornote{Corresponding author.}\email{lukeo@illinois.edu}\orcid{0000-0002-5283-6104}
\affiliation{%
  \institution{University of Illinois Urbana--Champaign}
  \city{Urbana}
  \state{Illinois}
  \country{USA}
}

\renewcommand{\shortauthors}{Diener et al.}

\begin{abstract}
\mirge{} is a computational approach for scientific computing based on
  \software{NumPy}-like array computation, but using lazy evaluation to recast
computation as data-flow graphs, where nodes represent immutable, multi-dimensional arrays.
Evaluation of an array expression is deferred until its value is needed, at which point a pipeline
is invoked that transforms high-level array expressions into lower-level intermediate
representations (IR) and finally into executable code, through a multi-stage process.
Domain-specific transformations, such as metadata-driven optimizations, GPU-parallelization
strategies, and loop fusion techniques, improve performance and memory efficiency.

\mirge{} employs ``array contexts'' to abstract the interface
between array expressions and heterogeneous execution environments (for example, lazy
evaluation via OpenCL, or eager evaluation via \software{NumPy} or \software{CuPy}).
The framework thus enables performance portability as well as separation of concerns between
application logic, low-level implementation, and optimizations.
By enabling scientific expressivity while facilitating performance tuning, \mirge{} offers a
robust, extensible platform for both computational research and scientific
application development.

This paper provides an overview of \mirge{}.
We further describe an application of \mirge{} called \mirgecom{}, for supersonic combusting
flows in a discontinuous Galerkin finite-element setting.
We demonstrate its capabilities as a solver and highlight its performance
characteristics on large-scale GPU hardware.
\end{abstract}


\received{X}
\received[revised]{X}
\received[accepted]{X}

\maketitle

\section{Introduction}

The design of scientific simulation software often involves making uncomfortable trade-offs between high-level
mathematical abstractions that facilitate understandability and low-level implementation details for
performance.
The latter can lead to complex codes that make maintenance and portability difficult.
In this paper, we present \mirge{} (\textbf{M}ath--\textbf{I}ntermediate
\textbf{R}epresentation--\textbf{G}eneration--\textbf{E}xecution), an approach that leverages the
high-level usability of Python and targets performance portability across modern and anticipated
accelerator architectures.

At the core of \mirge{} is a philosophy of \emph{separation of concerns}: the
scientific application is written using intuitive array expressions
equivalent to \software{NumPy} expressions in Python, while transformations and
optimizations are used to achieve performance. Crucially, these transformations
are application-specific (though with reusable parts) and driven by the user
through array context customization. This design enables
rapid simulation development without precluding the use of performance techniques such
as GPU parallelization and loop fusion.

Scientific codes are challenging due to the range of demands, including
\begin{enumerate*}[label=(\arabic*)]
\item readability for both domain experts and developers,
\item the need to attain (and maintain) performance on modern platforms,
\item ease of maintenance and extension, and
\item identifying performance opportunities and challenges via modeling.
\end{enumerate*}
One tenet of our strategy is that the clarity of physical and model
expressions should not be compromised by implementation details,
hardware-specific considerations, or performance optimizations.

\mirge{} is an approach that forms the backbone for the implementation of the \mirgecom~simulation
library, a solver for high-speed combustion.
This approach allows a straightforward description of a computational model
that closely resembles the mathematics of the physical model without
prescribing platform-specific algorithms that will be used to achieve performance.

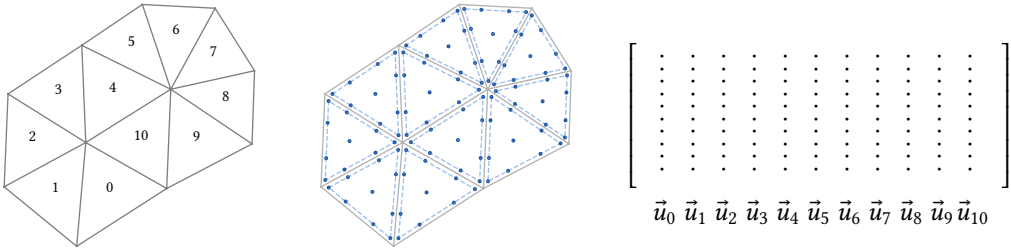
\begin{figure}[!ht]
  \centering
\begin{tikzpicture}[x=20pt,y=20pt]   %
  \node[anchor=south west] at (0,0) {\includegraphics{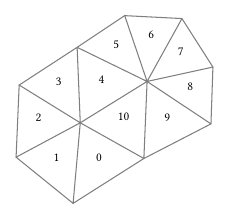}};
  \node[anchor=south west] at (6,0) {\includegraphics{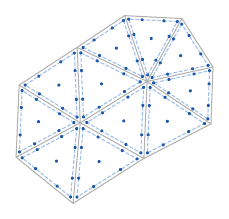}};
  \begin{scope}[shift={(320pt, 60pt)}]
    \matrix (A) [matrix of math nodes,left delimiter={[},right delimiter={]},text width=5pt]
    {
      \cdot & \cdot & \cdot & \cdot & \cdot & \cdot & \cdot & \cdot & \cdot & \cdot & \cdot \\[-5pt]
      \cdot & \cdot & \cdot & \cdot & \cdot & \cdot & \cdot & \cdot & \cdot & \cdot & \cdot \\[-5pt]
      \cdot & \cdot & \cdot & \cdot & \cdot & \cdot & \cdot & \cdot & \cdot & \cdot & \cdot \\[-5pt]
      \cdot & \cdot & \cdot & \cdot & \cdot & \cdot & \cdot & \cdot & \cdot & \cdot & \cdot \\[-5pt]
      \cdot & \cdot & \cdot & \cdot & \cdot & \cdot & \cdot & \cdot & \cdot & \cdot & \cdot \\[-5pt]
      \cdot & \cdot & \cdot & \cdot & \cdot & \cdot & \cdot & \cdot & \cdot & \cdot & \cdot \\[-5pt]
      \cdot & \cdot & \cdot & \cdot & \cdot & \cdot & \cdot & \cdot & \cdot & \cdot & \cdot \\[-5pt]
      \cdot & \cdot & \cdot & \cdot & \cdot & \cdot & \cdot & \cdot & \cdot & \cdot & \cdot \\[-5pt]
      \cdot & \cdot & \cdot & \cdot & \cdot & \cdot & \cdot & \cdot & \cdot & \cdot & \cdot \\[-5pt]
      \cdot & \cdot & \cdot & \cdot & \cdot & \cdot & \cdot & \cdot & \cdot & \cdot & \cdot \\[-5pt]
      };
      \foreach \x in {1,2,...,11}
        \pgfmathtruncatemacro{\z}{\x-1}
        \node[yshift=-10pt] at (A-10-\x.south) {$\vec{u}_{\z}$};
  \end{scope}
\end{tikzpicture}

  \Description{An example triangular mesh with 11 elements and 10 nodes marked within each element.  The corresponding 10 by 11 array is also shown as a matrix.}
  \caption{Discontinuous Galerkin layout.  A mesh (left) with nodal degrees-of-freedom local to each mesh element (middle), leads to an array-based layout of the unknowns (right).}\label{fig:dg_mesh}
\end{figure}

In a way, \mirge{} takes a similar approach to \software{JAX}~\cite{jax} and
\software{torch.fx}~\cite{torchfx}.  \software{JAX}, for example, uses what it calls
\textit{tracing} to record the operations of a function.
It then reconstructs a potentially more efficient function call for execution on
devices\footnote{https://docs.jax.dev/en/latest/jit-compilation.html}.
\software{torch.fx} has a focus on
capturing \software{PyTorch} code and transformations for device execution,
but has a clear focus ``on the DAG representation of deep learning programs.''
Similar \textit{graph mode} execution is pursued by a number of codes;
\mirge{} has a focus
on array computations, and consequently PDE-focused numerical simulation. One
reason for this is provided by~\cref{fig:dg_mesh}.  On the left is an example
triangular mesh (other meshes may also be used, such as 3D tetrahedral, or
quadrilateral and hexahedral meshes which may provide benefits when using a
tensor-product basis).  In a nodal discontinuous Galerkin
setting~\cite{HesthavenWarburton_dg_2007}, degrees of freedom (DOFs) are local to an
element (as viewed in the middle figure), which has a natural representation
in an array format (right figure).  As a result, the numerical methods
generally operate on arrays of data in bulk, providing an avenue to
organize the computation for overall efficiency.

\mirge{} is one of many approaches to providing user-level productivity (in
Python) while pursuing computational efficiency under the hood. \software{CuPy},
for example, is Python-based with a focus on easily executing
\software{NumPy}-like code on the GPU, providing both drop-in \software{NumPy}
replacements as well as an interface to write kernels directly. Likewise,
\software{PyOpenCL} and \software{PyCUDA}~\cite{kloeckner_2012_pyopencl} offer
direct access to GPU execution
from Python, but offer less drop-in functionality.
In this respect, \mirge{} has more in common with \software{CuPy}, but it differs in its focus on
providing access to intermediate representations (in part provided by the
\software{loopy}~\cite{loopy} package) in order to facilitate code transformations.
Similar transformation-enabling approaches have been developed, for example in
\software{Hercules}~\cite{hercules}, which focuses on transformations into parallel code.
Their approach concentrates on the IR and transformations, but purposely steps away from existing
high-level programming languages such as Python. In contrast, \mirge{} strives to make such
optimization work more accessible by realizing it as a transformation step within a Python-based
processing chain.
A further point of differentiation is that the design of \mirge{} is not tied to a particular
execution target (AMD, NVIDIA, CPUs, etc.); different backends may be used, enabling execution on
a variety of platforms.

In the next section, we take an in-depth look at past and present work, broadly categorizing
existing approaches and giving a sense of how \mirge{} fits among them.
Subsequent sections provide an overview of \mirge{}, detailing the lazy/\software{JAX}-style
evaluation for general array-based computation, achieving the overarching goal
of benefiting from Python-level semantics with low-level performance.  A
high-level description of \mirge{} is given in~\cref{sec:mirgeoverview},
setting the foundation and providing additional motivations.  The details of
\mirge{} are given in~\cref{sec:mirgedetails}, where we delve into the
intricacies of tracing the array-based code, generating intermediate
representations for further optimizations, and ultimately generating
code for execution on a variety of platforms.
In~\cref{sec:applications}, \mirgecom{} is introduced, an application of
\mirge{} for supersonic combusting flows.  We provide performance measurements on baseline
applications, show comparisons with existing codes, and demonstrate the use of
\mirgecom{} in practice.

\section{Related Work}

Array programming as a paradigm has been studied extensively.  Early systems such as
\software{Fortran-I}~\cite{backus1978fortran} introduced multidimensional arrays
primarily as a storage format. Iverson et al.~\cite{iverson_1978_apl_evolve}
designed \software{APL}, which adopted notation from tensor analysis to represent
$n$-dimensional arrays. Every object in \software{APL} is an array, and the language supports
high-level routines such as reshaping, rank querying, and inner or
outer-products. Successors to \software{APL} include \software{J}, \software{K},
\software{MATLAB}, \software{NumPy}~\cite{harris2020array}, and
\software{Julia}~\cite{bezanson2020julia}, all of which have contributed to the
popularization of array-centric computing for a wide range of scientific
applications.

Two key factors influence the usability of an array programming system for
computational science workloads. The first is its \emph{execution throughput}, which
determines the scale of problems that can be solved within a given amount of
compute time. The second is the \emph{debuggability} of the system, which governs the speed at
which computational scientists can program their workloads.  Improved
debuggability enables inspection of intermediate program states, thereby
accelerating the development process.  Both characteristics are direct
consequences of the system's execution model.

Toolkits like \software{NumPy}, early versions of
\software{Torch}~\cite{chintala2019torch},
\software{CuPy}~\cite{cupy_learningsys2017}, \software{APL}, and
\software{ArrayFire}~\cite{malcolm_2012_arrayfire} employ an \emph{eager
evaluation} model. In this model, each operation is executed as soon as it is
encountered, with its result persisted in memory. This allows progressive
evaluation~\cite{green_1996_cognitive_dims} of the program which aids in rapid
prototyping, but incurs performance overheads due to frequent memory roundtrips
for the operands and, in a GPU setting, repeated kernel launches. An alternative is the
\emph{deferred evaluation} programming model, where a computational graph is
constructed from the program, which is subsequently lowered by a compiler and
executed through a runtime layer. These frameworks address much of the
previously covered performance limitations, by techniques such as kernel fusion,
and, using tuned micro-kernel libraries such as \software{BLAS}~\cite{blas},
\software{cuDNN}~\cite{chetlur2014cudnn}. Examples of such systems include,
\software{Tensorflow}~\cite{abadi_2016_tensorflow}, \software{JAX}~\cite{jax2},
\software{Julia}~\cite{bezanson2020julia}, \software{Torch.fx}~\cite{torchfx},
\software{Theano}~\cite{thetheanodevelopmentteam2016theano}, and
\software{Legate}~\cite{bauer2019legatenumpy}. The runtime layers construct an
IR corresponding to the user's program either via tracing~\cite{jax2}, or
abstract syntax tree (AST) traversal~\cite{kwan_2015_numba} or a combination of
both~\cite{Thakkar_CUTLASS_2023}. The IR corresponding to the user's program is
subsequently leveraged by optimizing compilers.

By design, deferred evaluation is amenable to more sophisticated
optimizations than eager evaluation because the compiler has additional
information about the users of intermediate array operations. This information
is crucial for many compiler techniques, including kernel
fusion~\cite{kristensen_2016_bohirum_opt}, array
contraction~\cite{gao_1993_array_contract}, and data-layout transformations.
Transforming an array graph into a computational kernel that achieves
near-roofline performance is challenging primarily because of the large
transformation space, which includes the selection of execution-grid sizes,
decisions about materialization, the choice of address spaces, and more. Prior
work has shown that many subproblems within the lowering process are NP-hard.
For example, Rosenkrantz et al.~\cite{rosenkrantz_2006_materializations}
established that deciding which arrays to materialize is NP-complete; Lam et
al.~\cite{lam_1997_contractions} demonstrated that choosing the optimal
contraction path for Einstein summations is NP-hard; and Kennedy et
al.~\cite{kennedy_1994_loopfusion} showed that loop fusion for maximizing data
reuse is also NP-hard.  Consequently, many array frameworks adopt a pragmatic
choice by restricting the scope of compiler optimizations to those applicable
to a particular domain.  This trend is evident in frameworks such as
\software{TensorFlow}, \software{Jax}, and, \software{torch.fx}, which are primarily
used in machine-learning applications since the profitability of their  embedded
compiler optimizations does not transfer to other domains.

Multidimensional arrays are a fundamental data-structure widely seen across
various fields of scientific computing. However, we are not aware of any
state-of-the-art simulation engine that natively uses any of the array
frameworks mentioned above. Representative examples include
\software{Nek5000}~\cite{fischer2008nek5000,fischer_2022_nekrs}, which is a
\software{Fortran}-based Finite Element Methods (FEM) solver that provides
FEM-specific high-level routines for defining the weak-form.  Abdelfattah et
al.~\cite{abdelfattah2021nekrsperf} further optimized it for high-performance
architectures.  \software{MFEM}~\cite{mfem} is a \CPP{} FEM library that integrates
optimized assembly kernels from \software{libCEED}~\cite{brown2021libceed} and
leverages iterative solvers from \software{PETSc}~\cite{petsc-user-ref}.  The
Unified Form Language (UFL)~\cite{alnaes_2014_ufl} provides a high-level
abstraction for weak formulations in the Ciarlet
framework~\cite{ciarlet_2002_fem}, serving as the frontend for end-to-end
solvers such as \software{FEniCS}~\cite{alnaes_2015_fenics},
\software{Firedrake}~\cite{rathgeber_2017_firedrake}, and
\software{Dune}~\cite{bastian2021dune}.  These systems compile user-defined
variational forms into efficient kernels through FEM-specific
optimizations~\cite{kirby_2006_ffc,homolya_2018_tsfc,kulkarni_2025_fdrakegpu}.

Despite these advances, integrating user-defined array expressions into such
solvers remains impractical due to significantly low performance relative to the
roofline.  Prior efforts have partially addressed these concerns.
Hirata~\cite{hirata_2003_tce} proposed \software{Tensor Contraction Engine}, which provided
high-level operations for electronic structure computations, and allowed for
user-specified computations that were a subset of the tensor contraction
operations using the einsum notation.  Similarly,
\software{Comet}~\cite{gocken_2021_comet} allows computational chemistry programs
to be expressed in a subset of \software{NumPy}, lowering computations through the
tensor algebra dialect of \software{MLIR}~\cite{lattner_2021_mlir} and applying
optimizations such as Transpose-transpose-GEMM-transpose~\cite{hirata_2003_tce}.

In this work, we present a framework for scientific simulation development, together with an
application to a discontinuous Galerkin (DG) FEM solver, in which arrays are the sole
user-facing data-type.  Users can express computations entirely
through high-level array operations encompassing a pure subset of \software{NumPy}, i.e.,
a set of operations that disallows in-place modification.
We lower the resulting array graph through our \software{Pytato}/\software{Loopy} IR to
address a key limitation observed in prior work: the inability to inject
domain-specific code transformations.

\section{The \mirge{} Approach}\label{sec:mirgeoverview}

\begin{figure}[!thp]
    \includegraphics[width=\textwidth]{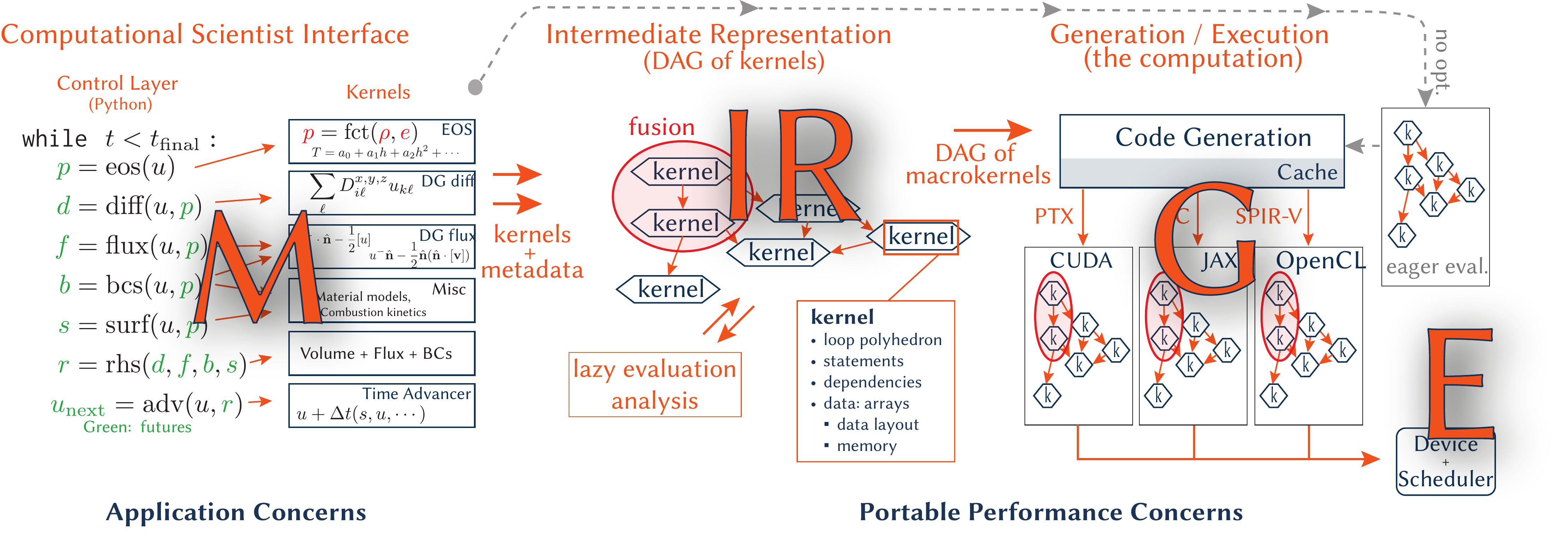}
    \Description{A graphic showing the \mirge{} process as a cartoon.  From left to right, the computational science interface is shown as a Python code listing followed by a collections of boxes that represent kernels.  These kernels map into a code generation box, where different code is output (for example, OpenCL or CUDA).}
    \caption{The \mirge~approach to scientific computing. (Left)
    the domain-specific math,  (middle) the optimized,
    host-specific computation kernels, and (right) generation and execution of platform specific code.}\label{fig:mirge-control}
\end{figure}

\Cref{fig:mirge-control} illustrates the \mirge~approach, from domain-specific
math (left) to transformed, host-specific computational kernels (middle) to code generation and
execution on target platforms (GPUs or CPUs). The non-optimized eager path, shown as a dashed grey line
(top), is a fall-back approach outside the intended \emph{lazy} processing path
for execution. This approach is novel in predictive science and provides a
two-way street between performance research and scientific simulation by
allowing access to large-scale predictive science applications and by
delivering the performance benefits directly to the application.

\begin{figure}[!thp]
\begin{tikzpicture}[x=1pt,y=1pt]
    \node[text width=120pt,anchor=north west] (a) at (0, 300)  {
\begin{lstlisting}[language=Python, basicstyle=\ttfamily\tiny]
import numpy as np
from arraycontext import ArrayContext

actx = ArrayContext()

def axpy_max0(x, y, a):
    w = actx.np.multiply(x, a)
    w = actx.np.add(w, y)
    w = actx.np.maximum(w, 0)
    return w

x = np.linspace(0, 1, 10)
y = np.linspace(0, 1, 10)
a = 0.5

axpy_max0_compiled = actx.compile(axpy_max0)
axpy_max0_compiled(x, y, a)
\end{lstlisting}
    };
\node[signal,fill=black, signal to=east,signal from=west,text=white,anchor=west, rotate=-10,font=\small] at (140,280) {Stage 1};
\node[signal,fill=black, signal to=east,signal from=west,text=white,anchor=west, rotate=-10,font=\small] at (210,190) {Stages 2--5};
  \begin{scope}[x=12pt, y=12pt,
    every node/.style={draw, circle, minimum size=15pt, font=\tiny, inner sep=0pt},
    arrow/.style={Latex-},
    xshift=190pt,
    yshift=200pt,
    anchor=center,
    ]
\node (plus) at (0,0) {\color{tikzred}\code{max}};

\node (einsum2) at (2.5,2) {\color{tikzred}$+$};

\node (dx) at (-4,4) {0};

\node[dashed] (L) at (1,4) {\color{tikzblue}$y$};
\node (index) at (4,4) {\color{tikzred}$*$};

\node[dashed] (i) at (3,6) {\color{tikzblue}$a$};
\node[dashed] (dof2) at (5,6) {\color{tikzblue}$x$};

\draw[arrow] (plus) -- (einsum2);

\draw[arrow] (plus) -- (dx);

\draw[arrow] (einsum2) -- (L);
\draw[arrow] (einsum2) -- (index);

\draw[arrow] (index) -- (i);
\draw[arrow] (index) -- (dof2);
\end{scope}

    \node[text width=120pt,anchor=north east] (a) at (395, 300)  {
\begin{lstlisting}[language=C, basicstyle=\ttfamily\tiny]
// OpenCL kernel version

__kernel void axpy_max0(
    __global const float *x,
    __global const float *y,
    const float a,
    __global float *w,
    const int n)
{

    int i = get_global_id(0);

    if (i < n) {
        float t = a * x[i] + y[i];
        w[i] = t > 0.0f ? t : 0.0f;
    }

}
\end{lstlisting}
  };
\end{tikzpicture}
\Description{A Python code listing on the left follows into a graph followed by an OpenCL code listing.}
\caption{Stages 1 and 2--5 of the \mirge{} approach for an example operation on two arrays $x$ and $y$ (with scalar $a$): $\max(0, y+ax)$.
(Left) \software{NumPy}-like application code using the \mirge{} approach.
(Middle) The array dataflow graph (ADFG), with the nodes representing array-valued expressions, while the edges represent dependencies between them.
(Right) Generated OpenCL code after the transformations in the previous.}\label{fig:code-example-all}
\end{figure}
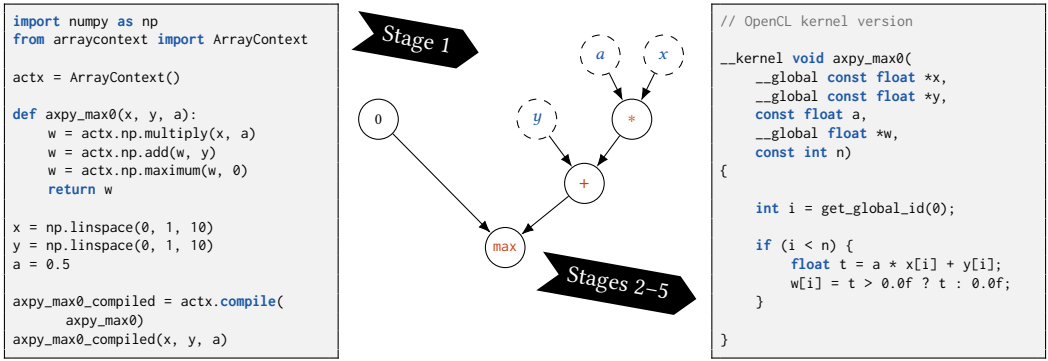

We make use of a pipeline that uses lazy execution to enable code transformations.
This relies on a user-facing layer of \software{NumPy}-like array constructs (an \emph{array context})
that internally captures computation symbolically, deferring execution to enable performance
optimizations and to target different execution environments, such as GPUs.
Arrays in this framework are considered immutable objects that represent the fundamental
building blocks of the computation.
By evaluating the program in this manner and producing an \emph{array dataflow graph} (ADFG),
a comprehensive view of the user's intended computation is available and transformations such as fusion and
parallelization can be applied. An example of the user-facing part of this is shown on the left side of
\cref{fig:code-example-all}. Here, array operations
in \code{axpy\_max0()} are evaluated symbolically and code is generated from the resulting ADFG,
which is then executed.

How does the simulation unfold?  Over several stages the steps of the
computation are collected, transformed into a target language (e.g. OpenCL),
and executed:
\begin{description}
  \item[Stage 1: Capture an Array Dataflow Graph (ADFG)]

The pipeline begins by representing the mathematical steps of computation as an array dataflow graph
(ADFG), where each node represents an array or array operation.
    The calculation, represented as a \software{NumPy}-like array program, is traced symbolically to
assemble the ADFG.
Tracing proceeds until the execution reaches a user-defined stopping point (via \code{actx.freeze()}/
\code{actx.compile()}), at which point the rest of the code generation pipeline is triggered.
\Cref{fig:code-example-all} shows the ADFG in the middle;
nodes represent array-valued expressions, while the edges represent
dependencies between them.

\item[Stage 2: Transform the ADFG]

Next, domain-specific transformations are applied to the ADFG, such as algebraic simplifications
and an initial decision on materialization, that is, whether any given intermediate value is stored in memory. Much of this process is driven by sparse, descriptive metadata that is attached to the ADFG nodes in Stage~1 by user or library code.

\item[Stage 3: Rewrite to a Scalar Intermediate Representation (IR)]

Next, the transformed ADFG is converted into a scalar (loop-level) intermediate
representation (IR), which introduces loop structures and memory allocations.
Metadata may also propagate into the IR in the next stage, as
metadata on loop variables, for example.
While the ADFG expresses \emph{what} should be computed, the IR expresses \emph{how}
it should be computed, including execution order and memory ordering.

\item[Stage 4: Transform the Scalar IR]

With this IR, transformations for parallelization and memory access are applied.
For example, it is often advantageous for loops to be partitioned into block/thread structures.  Additionally, metadata, propagated
from prior stages, may be used to determine which loops to fuse and how to map array axes
onto parallel hardware.
This stage is home to much of the performance-tuning work, and is an opportunity to link with
external tools.

\item[Stage 5: Emit Generated Target Code]

  After optimization transformations, the IR is used to generate target code.  Our examples use OpenCL, though \software{JAX} or CUDA are other options.
This stage also translates higher-level constructs into efficient low-level code, potentially
relying on OpenCL-specific code for memory access and parallel execution.

\item[Stage 6: Program Execution]

In the final stage, the compiled kernel is executed on the target hardware. The array
context manages the memory allocation, the OpenCL execution context, and the
launching of kernels.  Since earlier stages have already optimized and compiled the program, the
runtime overhead is minimal.

\end{description}

We make use of two separate intermediate representations in our transform chain to focus
on distinct aspects: The ADFG for the most part does not represent `space' (memory) and `time'
(evaluation order and parallelization), and thus transformations applied at its level are mostly
of an algebraic nature. Meanwhile, optimizations that depend on these specifics are applied
at the level of the scalar/loop-level IR.

\section{Stages of a Computation: Finer Points}\label{sec:mirgedetails}

The previous section provided a high-level overview of the overall approach to
\mirge{}.  Next, we examine the details, highlighting necessary
pieces for generality and for achieving performance at scale.

As previously mentioned, array operations in \mirge{} are expressed using \software{NumPy}-like
\emph{array contexts}. Nominally, array contexts control the execution of the array
programs supplied by the user. More specifically, this involves choice of the execution
environment and control of program transformation for high performance. To this end, the
array programming interface provided is focused on facilitating lazy evaluation, though
other execution environments are also available, cf. \cref{tab:array-contexts}.
\begin{table}[!thp]
\caption{Overview of eager and lazy evaluation array contexts. The \textbf{Lazy PyOpenCL-based} array context is described in detail in this paper.}\label{tab:array-contexts}
\begin{tabular}{ll}
\toprule
Eager evaluation array contexts & Lazy evaluation array contexts\\
\midrule
Eager PyOpenCL-based & \textbf{Lazy PyOpenCL-based} \\
  \software{JAX}-eager & \software{JAX}-lazy \\
  \software{NumPy} & \\
  \software{CuPy} & \\
\bottomrule
\end{tabular}
\end{table}

\begin{figure}[!thp]
\includegraphics{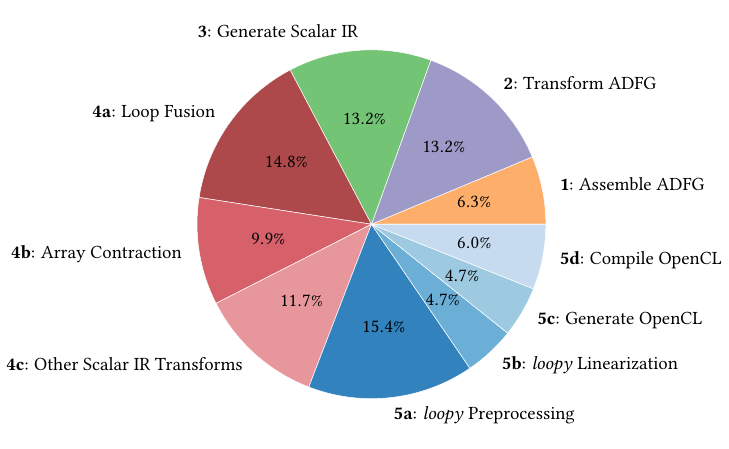}
\Description{A pie chart with stage 1 Assemble ADFG representing 6.3 percent, stage 2 Transform ADFG as 13.2 percent, stage 3 Generate Scalar IR as 13.2 percent, stages 4a Loop Fusion, 4b Array Contraction, and 4c Other Scalar IR Transforms as 14.8, 9.9, and 11.7 percent, and stages 5a loopy Preprocessing, 5b loopy Linearization, 5c Generate OpenCL, and 5d Compile OpenCL as 15.4, 4.7, 4.7, and 6.0 percent.}
\caption{Breakdown of time spent in pre-execution stages for an example case.}\label{fig:compile-time-chart}
\end{figure}

To give the reader a sense of the complexity of processing applied at the various stages of the transform chain,
\Cref{fig:compile-time-chart} shows a breakdown of the fraction of time spent at each stage
for an example problem (a compressible multispecies flow case similar to the case
presented in \cref{sec:ceesd-prediction}, with a smaller mesh of approximately 200K degrees of freedom).
In our applications, the processing time has little to no dependence on the sizes of the array data.
There is some dependence on the layout of data across processors in distributed
environments, but this is bounded: it grows with the number of adjacent neighbor processors, which
eventually saturates.
Instead, scaling has a complex dependence on the amount and nature of the work specified in the
application code.
Stages 1 and 2 generally scale with the number of nodes in the ADFG, whereas the latter stages
introduce additional scaling dependencies, such as the number of generated statements and the
relationships between them.
In practice, the illustrated chart is qualitatively representative of what we observe for most problems.
Thus, our primary approach to managing processing time has been to employ strategies to minimize the
size of the initial ADFG (see \cref{sec:outlining} for an example).

\subsection{Stage 1: Capture an Array Dataflow Graph (ADFG)}

The first stage of the compilation pipeline begins by capturing a symbolic representation of the
computation, i.e., the ADFG.
Various node types are used to represent the operations; \cref{tab:adfg-nodes} summarizes some of
the most common types.

\begin{table}[!thp]
\caption{A selection of node types commonly found in ADFGs.}\label{tab:adfg-nodes}
\begin{tabular}{L{7cm}M{3.5cm}}
\toprule
\textbf{Node type} & \textbf{Example(s)} \\
\midrule
\textbf{Data} \newline Constant, numerical-valued data arrays &
\begin{tikzpicture}
\small
\node[draw, circle, minimum width=0.7cm, minimum height=0.7cm] {$a$};
\end{tikzpicture} \\[10pt]
\textbf{Placeholders} \newline Yet-to-be-supplied input arrays &
\begin{tikzpicture}
\small
\node[draw, dashed, circle, minimum width=0.7cm, minimum height=0.7cm] {\color{tikzblue}$x$};
\end{tikzpicture} \\[10pt]
\textbf{Index lambdas} \newline Elementwise algebraic operations &
\begin{tikzpicture}
\small
\node[draw, circle, minimum width=0.7cm, minimum height=0.7cm] at (-0.5,0) {\color{tikzred}$+$};
\node[draw, circle, minimum width=0.7cm, minimum height=0.7cm] at (0.5,0) {\color{tikzred}$*$};
\end{tikzpicture} \\[10pt]
\textbf{Reshapes} \newline Changes to logical array shape &
\begin{tikzpicture}
\small
\node[draw, ellipse, minimum width=0.8cm, minimum height=0.8cm, inner sep=0] {\color{tikzred}$\rightarrow$\code{(3,2)}};
\end{tikzpicture} \\[10pt]
\textbf{Indexing} \newline Array slicing and indirect indexing &
\begin{tikzpicture}
\small
\node[draw, ellipse, minimum width=0.7cm, minimum height=0.7cm] {\color{tikzred}\code{[:,0]}};
\end{tikzpicture} \\[10pt]
\textbf{Einsums} \newline Einstein-summation-like operations &
\begin{tikzpicture}
\small
\node[draw, ellipse, minimum width=0.8cm, minimum height=0.8cm, inner sep=0] {\color{tikzred}$ij,jk \rightarrow ik$};
\end{tikzpicture} \\[10pt]
\textbf{Joins} \newline Concatenation, stacking, etc. &
\begin{tikzpicture}
\small
\node[draw, ellipse, minimum width=1.0cm, minimum height=0.7cm] (concat) at (-1.1,0) {};
\node[draw=none] at ($(concat.center) + (-0.3,0)$) {\color{tikzred}$\square$};
\node[draw=none] at ($(concat.center) + (-0.075,0)$) {\color{tikzred}$\square$};
\node[draw=none] at ($(concat.center) + (0.275,-0.05)$) {\color{tikzred}$\cdots$};
\node[draw, ellipse, minimum width=0.7cm, minimum height=0.7cm, inner sep=0] at (0.7,0) {\color{tikzred}$(\square\hspace{0.05cm},\square\hspace{0.05cm},\cdots)$};
\end{tikzpicture} \\[10pt]
\textbf{Functions} \newline ``Outlined'' function definitions and calls &
\begin{tikzpicture}
\small
\node[draw, rectangle, minimum width=0.7cm, minimum height=0.7cm] at (-0.55,0) {\color{tikzgreen}\code{f}};
\node[draw, ellipse, minimum width=0.7cm, minimum height=0.7cm] at (0.85,0) {\color{tikzred}\code{f(}{\color{black}\ldots}\code{)}};
\end{tikzpicture} \\[10pt]
\textbf{Communication} \newline Distributed sends and receives &
\begin{tikzpicture}
\small
\node[draw, circle, minimum width=0.7cm, minimum height=0.7cm] at (-0.5,0) {\color{tikzgreen}$S$};
\node[draw, ellipse, minimum width=0.7cm, minimum height=0.7cm] at (0.7,0) {\color{tikzred}$R[{\color{black}a}]$};
\end{tikzpicture} \\
\bottomrule
\end{tabular}
\end{table}

The design of this array-level intermediate representation is intentionally constrained: it is
expressive enough to support transformations such as fusion, but restricted enough to limit the
complexity of implementing such transformations.
Unlike other IRs such as LLVM IR or MLIR, \mirge{}'s ADFG does not use indirection (e.g., name
strings) when referring to nodes, but rather direct object references.
This means that the ADFG is more expensive to modify, as it requires graph traversals for each
transformation, but on the other hand it does not require maintaining a global state or name mapping.
Together with the immutability of the array node data structure, this design choice simplifies the notion of equality
between graphs and makes it easier to find redundant computation.

\subsubsection{Compilation and Freezing}

As operating on lazy arrays simply appends nodes to the ADFG, intervention is required to trigger
code generation and execution. Early approaches to this issue (e.g., \cite{bohrium}) tended to trigger
code generation implicitly, such as when array entries were retrieved by the user.
Unfortunately, doing so incurs high and repeated costs rooted in building and processing of ADFGs,
at least to a point where existing code can be retrieved from cache.
In \mirge{}, we thus pursue a more explicit approach, with two interfaces available to the user:
If the code is intended to be called once, the user evaluates their array code and passes it to
\code{actx.freeze()} to trigger the rest of the pipeline, generate and ultimately execute the code.
If the code is intended to be executed multiple times, the user instead passes a function to
\code{actx.compile()}, which evaluates it using placeholder inputs, triggers the rest of the pipeline,
and returns an object that can be used for repeated execution.
Examples of both of these are shown in \cref{fig:freeze-compile}.
\begin{figure}[!thp]
\centering
\begin{subfigure}[b]{0.45\textwidth}
\centering
\begin{lstlisting}[language=Python, basicstyle=\ttfamily\tiny]
a = actx.from_numpy(np.random.rand(5,2))

a_sum_sq = actx.np.sum(a**2)  # ADFG
a_sum_sq_frozen = actx.freeze(a_sum_sq)  # device array
a_sum_sq_np = actx.to_numpy(a_sum_sq_frozen)  # numpy array
\end{lstlisting}
\caption{Using \code{freeze()}.}
\end{subfigure}
\hspace{0.09\textwidth}%
\begin{subfigure}[b]{0.45\textwidth}
\centering
\begin{lstlisting}[language=Python, basicstyle=\ttfamily\tiny]
def sum_sq(x):
    return actx.np.sum(x**2)

sum_sq_compiled = actx.compile(sum_sq)

a = actx.from_numpy(np.random.rand(5,2))

a_sum_sq = sum_sq_compiled(a)  # device array
a_sum_sq_np = actx.to_numpy(a_sum_sq)  # numpy array
\end{lstlisting}
\caption{Using \code{compile()}.}
\end{subfigure}
  \Description{Python code showing a freeze of \software{NumPy} sum and Python code showing a compile of \software{NumPy} sum.}
\caption{Methods for triggering code generation/execution when using lazily-evaluated arrays.}\label{fig:freeze-compile}
\end{figure}

It is important to note that the implementation of \code{compile()} itself is also lazy.
At the time of calling \code{compile()}, arguments that will be passed to the compiled function are
unknown; in order to avoid burdening the user with the specification of argument metadata up front,
\code{compile()} defers most of its processing until the first time the compiled function is called.
This also allows the same compiled function to be called with different configurations of optional
arguments; separate code will be generated for each case, with reuse occuring when applicable.

\subsubsection{Distributed Computation}

In \mirge{}, array-context-based calculations may be distributed across multiple compute units.
Therefore, consideration must be given to the representation of ADFGs that result from lazy evaluation in
such environments.
One approach, analogous to a \emph{single control} stream model of computation (PGAS, distributed arrays),
is to construct a shared, global ADFG containing the entire calculation.
A na\"ive implementation of this, while workable at small scales, breaks down when constant array
data (such as connectivity data in a finite element mesh) becomes too large to fit in a single compute unit's memory.
More sophisticated variations may be able to avoid such problems by storing constant data
separately from nodes, but this increases the amount of bookkeeping required when applying
transformations to the ADFG.
Instead, we take the approach of \emph{multiple control streams}, constructing separate, local ADFGs for
each compute unit and embedding nodal representations of inter-process communication within them.
From the perspective of a user, this amounts to writing the application much like a typical MPI-based
code, but rather than calling out to MPI directly, special routines are called to create send and
receive nodes and to insert them into the ADFG.

One challenge to this approach is that sends are not arrays in a dataflow sense; they are
``dead end'' nodes without an output.
This makes them unsuitable for direct representation in an ADFG.
Presently, the \mirge{} infrastructure circumvents this issue by introducing a node type
that:
\begin{enumerate*}[label=(\arabic*)]
\item acts as a passthrough for an existing array in the ADFG, and
\item holds a reference to the data (as represented by an ADFG node) to be sent.
\end{enumerate*}
While this approach is viable, it is not without downsides.
Namely, one must ensure that the sub-expression containing the send is retained as a dependency
of the ADFG (in other words, its result is used in the overall computation); otherwise, unpaired
communication (i.e., a receive without a send or vice versa) may result.
Alternative methods, such as explicit tracking of sends in user APIs, are under
investigation.

\subsubsection{ADFG Functions}\label{sec:outlining}

Because the dataflow graphs are generated through symbolic execution of array code, function
calls are not automatically represented explicitly.
Instead, each function call creates a new subgraph of its operations, analogous to
inlining the code for each call.
As a result, the ADFG size can grow, leading to increased processing time.
Hence, it is beneficial to provide a means of selectively ``outlining'' functions~---~i.e., preventing
a function call from being inlined into the ADFG.

\begin{figure}[!thp]
  \centering
  \begin{subfigure}[t]{0.49\textwidth}
    \centering
    \begin{tikzpicture}[
      >=Latex,
      every node/.style={draw, circle, minimum width=0.7cm, minimum height=0.7cm},
      node distance=0.2cm and 0.2cm
    ]
    \small
    \node (a) at (-2.25,-0.6) {$a$};
    \node (b) at (-0.75,-0.6) {$b$};
    \node (c) at (0.75,-0.6) {$c$};
    \node (d) at (2.25,-0.6) {$d$};
    \node (asq) at ($(a)+(0,-1.5)$) {\color{tikzred}$*$};
    \node[draw=none, rectangle, anchor=west, minimum height=0] (asqlabel) at (asq.east) {
      \color{tikzgray}\tiny$=a^2$};
    \node (bsq) at ($(b)+(0,-1.5)$) {\color{tikzred}$*$};
    \node[draw=none, rectangle, anchor=west, minimum height=0] (bsqlabel) at (bsq.east) {
      \color{tikzgray}\tiny$=b^2$};
    \node (csq) at ($(c)+(0,-1.5)$) {\color{tikzred}$*$};
    \node[draw=none, rectangle, anchor=west, minimum height=0] (csqlabel) at (csq.east) {
      \color{tikzgray}\tiny$=c^2$};
    \node (dsq) at ($(d)+(0,-1.5)$) {\color{tikzred}$*$};
    \node[draw=none, rectangle, anchor=west, minimum height=0] (dsqlabel) at (dsq.east) {
      \color{tikzgray}\tiny$=d^2$};
    \node (asqplusbsq) at ($(asq)!0.5!(bsq)+(0,-1.5)$) {\color{tikzred}$+$};
    \node[draw=none, rectangle, anchor=west, minimum height=0] (asqplusbsqlabel) at (asqplusbsq.east) {
      \color{tikzgray}\tiny$=a^2 + b^2$};
    \node (csqplusdsq) at ($(csq)!0.5!(dsq)+(0,-1.5)$) {\color{tikzred}$+$};
    \node[draw=none, rectangle, anchor=west, minimum height=0] (csqplusdsqlabel) at (csqplusdsq.east) {
      \color{tikzgray}\tiny$=c^2 + d^2$};
    \node (out) at ($(asqplusbsq)!0.5!(csqplusdsq)+(0,-1.5)$) {\color{tikzred}$+$};
    \node[draw=none, rectangle, anchor=west, minimum height=0] (outlabel) at (out.east) {
      \color{tikzgray}\tiny$=a^2 + b^2 + c^2 + d^2$};
    \draw[->,bend left] (a) edge (asq);
    \draw[->,bend right] (a) edge (asq);
    \draw[->,bend left] (b) edge (bsq);
    \draw[->,bend right] (b) edge (bsq);
    \draw[->,bend left] (c) edge (csq);
    \draw[->,bend right] (c) edge (csq);
    \draw[->,bend left] (d) edge (dsq);
    \draw[->,bend right] (d) edge (dsq);
    \draw[->] (asq) -- (asqplusbsq);
    \draw[->] (bsq) -- (asqplusbsq);
    \draw[->] (csq) -- (csqplusdsq);
    \draw[->] (dsq) -- (csqplusdsq);
    \draw[->] (asqplusbsq) -- (out);
    \draw[->] (csqplusdsq) -- (out);
    \end{tikzpicture}
    \caption{Inlined}
  \end{subfigure}%
  ~
  \begin{subfigure}[t]{0.5\textwidth}
    \centering
    \begin{tikzpicture}[
      >=Latex,
      every node/.style={draw, circle, minimum width=0.7cm, minimum height=0.7cm},
      node distance=0.2cm and 0.2cm
    ]
    \small
    \node (a) at (-3,-0.6) {$a$};
    \node (b) at (-2, -0.6) {$b$};
    \node (c) at (2, -0.6) {$c$};
    \node (d) at (3, -0.6) {$d$};
    \node[ellipse] (callf1) at ($(a)!0.5!(b)+(-0,-1.5)$) {\footnotesize\color{tikzred}\code{f(}{\color{black}\ldots}\code{)}};
    \node[ellipse] (callf2) at ($(c)!0.5!(d)+(0,-1.5)$) {\footnotesize\color{tikzred}\code{f(}{\color{black}\ldots}\code{)}};
    \node (result1) at ($(callf1)+(0,-1.5)$) {\footnotesize\color{tikzred}\code{res0}};
    \node[draw=none, rectangle, anchor=north, minimum height=0] (result1label) at (result1.south) {
      \color{tikzgray}\tiny$=a^2 + b^2$};
    \node (result2) at ($(callf2)+(0,-1.5)$) {\footnotesize\color{tikzred}\code{res0}};
    \node[draw=none, rectangle, anchor=north, minimum height=0] (result2label) at (result2.south) {
      \color{tikzgray}\tiny$=c^2 + d^2$};
    \node (out) at ($(result1)!0.5!(result2)+(0,-1.5)$) {\color{tikzred}$+$};
    \node[draw=none, rectangle, anchor=west, minimum height=0] (outlabel) at (out.east) {
      \color{tikzgray}\tiny$=a^2 + b^2 + c^2 + d^2$};
    \node[rectangle, minimum width=2.5cm, minimum height=3.5cm] (f) at (0, -2.4) {\color{tikzgreen}\code{f}};
    \node[dashed] (x) at (-0.6, -1.15) {\color{tikzblue}$x$};
    \node[dashed] (y) at (0.6, -1.15) {\color{tikzblue}$y$};
    \node (xsq) at ($(x)+(0,-1.3)$) {\color{tikzred}$*$};
    \node (ysq) at ($(y)+(0,-1.3)$) {\color{tikzred}$*$};
    \node (xsqplusysq) at ($(xsq)!0.5!(ysq)+(0,-1.2)$) {\color{tikzred}$+$};
    \draw[->] (a) -- (callf1);
    \draw[->] (b) -- (callf1);
    \draw[->] (c) -- (callf2);
    \draw[->] (d) -- (callf2);
    \draw[->] (callf1) -- (result1);
    \draw[->] (callf2) -- (result2);
    \draw[->] (result1) -- (out);
    \draw[->] (result2) -- (out);
    \draw[dashed] (callf1) -- (f);
    \draw[dashed] (callf2) -- (f);
    \draw[->,bend left] (x) edge (xsq);
    \draw[->,bend right] (x) edge (xsq);
    \draw[->,bend left] (y) edge (ysq);
    \draw[->,bend right] (y) edge (ysq);
    \draw[->] (xsq) -- (xsqplusysq);
    \draw[->] (ysq) -- (xsqplusysq);
    \end{tikzpicture}
    \caption{Outlined}
  \end{subfigure}
  \Description{For inlining, a binary tree with three levels with $a$, $b$, $c$, and $d$ leading to a multiplication operator.  Each pair of multiplication leads to addition.  The two additions lead to another addition resulting in $a^2 + b^2 + c^2 + d^2$.  For outlining, a function $f$ is represented as a binary tree with inputs $x$ and $y$ leading to a multiplication operation each, which lead to an addition operation.  Then nodes $a$ and $b$ are sent to this function, and nodes $c$ and $d$ are sent to this function.  The results point to an addition operation again resulting in $a^2 + b^2 + c^2 + d^2$
.}
  \caption{
    An example of function outlining for an ADFG containing two calls to a function
    {\color{tikzgreen}\code{f}} that computes the sum of the squares of inputs {\color{tikzblue}$x$} and
    {\color{tikzblue}$y$}. Note: functions can have multiple results, hence call nodes
    ({\color{tikzred}\code{f(}\ldots\code{)}}) and result nodes (\code{\color{tikzred}res0}) are
    represented separately in the ADFG.
  }\label{fig:outlining}
\end{figure}
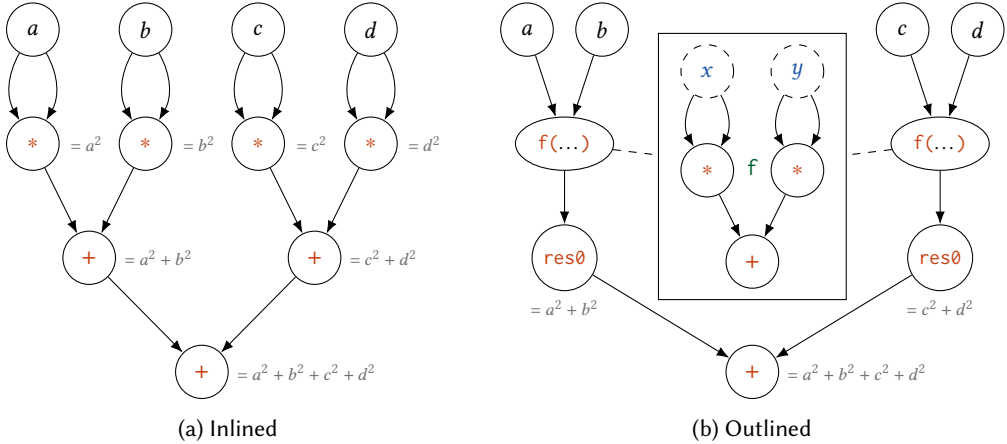

When a function is outlined, instead of directly returning the inlined nodes for the operations of
a function, a node representing the function call is returned.
This node contains a representation of the function's nodes that is isolated from the rest of the
ADFG by using placeholders for the function arguments.
The corresponding concrete arguments are also stored separately in the call node.
An example is shown in~\cref{fig:outlining}.

Representing function calls explicitly in the ADFG allows them to be processed by the downstream code
generation pipeline into kernel functions, or in certain cases to be optimized via ADFG
transformation. See~\cref{sec:concatenation}
for an example of this.

\subsection{Stage 2: Transform the Array DFG}

With an initial dataflow graph in place, the graph is parsed and transformed to
prepare for the later stages.  This includes identifying mathematical
simplifications in the graph, marking nodes for storage (or
\emph{materialization}), organizing the graph for distributed execution, and
concatenating certain function calls for efficiency.  We detail each of these
concepts next; as a whole they represent the ``transformation stage'' of the ADFG.

\subsubsection{Apply Math Simplifications}

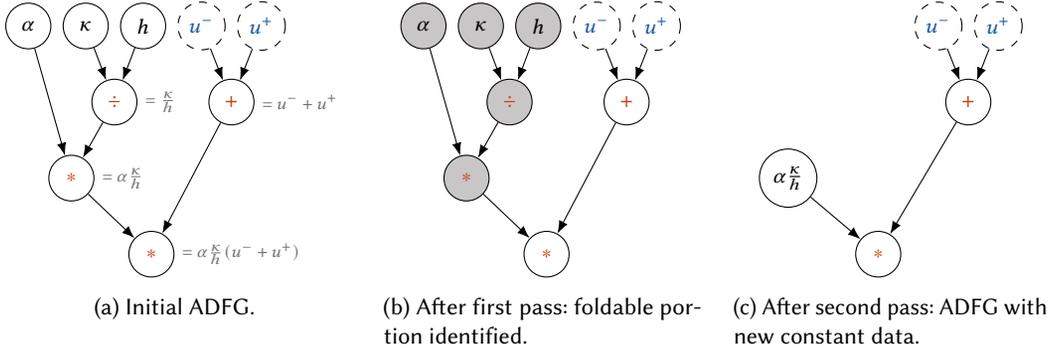
\begin{figure}[!thp]
  \centering
  \begin{subfigure}[t]{0.33\textwidth}
    \centering
    \begin{tikzpicture}[
      >=Latex,
      every node/.style={draw, circle, minimum width=0.6cm, minimum height=0.6cm},
      node distance=0.15cm and 0.15cm
    ]
    \footnotesize
    \node (h) at (0,-0.6) {$h$};
    \node[left=of h] (kappa) {$\kappa$};
    \node[left=of kappa] (alpha) {$\alpha$};
    \node[dashed, right=of h] (uminus) {\color{tikzblue}$u^-$};
    \node[dashed, right=of uminus] (uplus) {\color{tikzblue}$u^+$};
    \node (kappah) at ($(kappa)!0.5!(h)+(0,-1)$) {\color{tikzred}$\div$};
    \node[draw=none, rectangle, anchor=west, minimum height=0] (kappahlabel) at (kappah.east) {
      \color{tikzgray}\tiny$=\frac{\kappa}{h}$};
    \node (uplusu) at ($(uminus)!0.5!(uplus)+(0,-1)$) {\color{tikzred}$+$};
    \node[draw=none, rectangle, anchor=west, minimum height=0] (uplusulabel) at (uplusu.east) {
      \color{tikzgray}\tiny$=u^- + u^+$};
    \node (alphakappah) at ($(alpha)!0.5!(kappah)+(0,-1.5)$) {\color{tikzred}$*$};
    \node[draw=none, rectangle, anchor=west, minimum height=0] (alphakappahlabel) at (alphakappah.east) {
      \color{tikzgray}\tiny$=\alpha\frac{\kappa}{h}$};
    \node (result) at ($(uplusu)!0.5!(alphakappah)+(0,-1.5)$) {\color{tikzred}$*$};
    \node[draw=none, rectangle, anchor=west, minimum height=0] (resultlabel) at (result.east)  {
      \color{tikzgray}\tiny$=\alpha\frac{\kappa}{h}(u^- + u^+)$};
    \draw[->] (kappa) -- (kappah);
    \draw[->] (h) -- (kappah);
    \draw[->] (uminus) -- (uplusu);
    \draw[->] (uplus) -- (uplusu);
    \draw[->] (alpha) -- (alphakappah);
    \draw[->] (kappah) -- (alphakappah);
    \draw[->] (alphakappah) -- (result);
    \draw[->] (uplusu) -- (result);
    \end{tikzpicture}
    \caption{Initial ADFG.}
  \end{subfigure}%
  \hfill
  \begin{subfigure}[t]{0.3\textwidth}
    \centering
    \begin{tikzpicture}[
      >=Latex,
      every node/.style={draw, circle, minimum width=0.6cm, minimum height=0.6cm},
      node distance=0.15cm and 0.15cm
    ]
    \footnotesize
    \node[fill=tikzlightgray] (h) at (0,-0.6) {$h$};
    \node[left=of h, fill=tikzlightgray] (kappa) {$\kappa$};
    \node[left=of kappa, fill=tikzlightgray] (alpha) {$\alpha$};
    \node[dashed, right=of h] (uminus) {\color{tikzblue}$u^-$};
    \node[dashed, right=of uminus] (uplus) {\color{tikzblue}$u^+$};
    \node[fill=tikzlightgray] (kappah) at ($(kappa)!0.5!(h)+(0,-1)$) {\color{tikzred}$\div$};
    \node (uplusu) at ($(uminus)!0.5!(uplus)+(0,-1)$) {\color{tikzred}$+$};
    \node[fill=tikzlightgray] (alphakappah) at ($(alpha)!0.5!(kappah)+(0,-1.5)$) {\color{tikzred}$*$};
    \node (result) at ($(uplusu)!0.5!(alphakappah)+(0,-1.5)$) {\color{tikzred}$*$};
    \draw[->] (kappa) -- (kappah);
    \draw[->] (h) -- (kappah);
    \draw[->] (uminus) -- (uplusu);
    \draw[->] (uplus) -- (uplusu);
    \draw[->] (alpha) -- (alphakappah);
    \draw[->] (kappah) -- (alphakappah);
    \draw[->] (alphakappah) -- (result);
    \draw[->] (uplusu) -- (result);
    \end{tikzpicture}
    \caption{After first pass: foldable portion identified.}
  \end{subfigure}%
  \hfill
  \begin{subfigure}[t]{0.3\textwidth}
    \centering
    \begin{tikzpicture}[
      >=Latex,
      every node/.style={draw, circle, minimum width=0.6cm, minimum height=0.6cm},
      node distance=0.15cm and 0.15cm
    ]
    \footnotesize
    \node[dashed] (uminus) at (0.8,-0.6) {\color{tikzblue}$u^-$};
    \node[dashed, right=of uminus] (uplus) {\color{tikzblue}$u^+$};
    \node (uplusu) at ($(uminus)!0.5!(uplus)+(0,-1)$) {\color{tikzred}$+$};
    \node (alphakappah) at (-1.2,-2.6) {$\alpha\frac{\kappa}{h}$};
    \node (result) at ($(uplusu)!0.5!(alphakappah)+(0,-1.5)$) {\color{tikzred}$*$};
    \draw[->] (uminus) -- (uplusu);
    \draw[->] (uplus) -- (uplusu);
    \draw[->] (alphakappah) -- (result);
    \draw[->] (uplusu) -- (result);
    \end{tikzpicture}
    \caption{After second pass: ADFG with new constant data.}
  \end{subfigure}
  \Description{Three binary trees representing the stages of folding.  First is the binary tree representing the operation $\alpha \frac{\kappa}{h}(u^- + u^+)$.  Next the nodes of the tree representing $\alpha \frac{\kappa}{h}$ is highlighted.  Finally, these nodes are removed and replaced with a single node representing $\alpha \frac{\kappa}{h}$.}
  \caption{
    Constant folding for an expression that operates on a mixture of constant data
    ($\alpha$, $\kappa$, and $h$) and placeholders ({\color{tikzblue}$u^-$}, {\color{tikzblue}$u^+$}).
  }\label{fig:constantfold}
\end{figure}

Transformation tools in \mirge{} are used to apply mathematical simplifications to the
intermediate ADFG\@.
One example is \textit{constant folding}, where pieces of the
computation that depend only on constant data are evaluated ahead-of-time,
as illustrated in~\cref{fig:constantfold}.

\subsubsection{Materialization}
\label{sec:materialization}

\begin{figure}[!thp]
  \centering
  \begin{subfigure}[t]{0.45\textwidth}
    \centering
    \begin{minipage}{0.45\textwidth}
      \centering
      \begin{tikzpicture}[
        >=Latex,
        every node/.style={draw, circle, minimum width=0.6cm, minimum height=0.6cm},
        node distance=0.15cm and 0.15cm
      ]
      \footnotesize
      \node[fill=tikzlightgray] (a) at (-1.1,0) {$a$};
      \node[fill=tikzlightgray] (x) at (0,0) {$x$};
      \node[fill=tikzlightgray] (y) at (1.1,0) {$y$};
      \node (ax) at ($(a)!0.5!(x)+(0,-1)$) {\color{tikzred}$*$};
      \node (axpy) at ($(x)+(0,-2)$) {\color{tikzred}$+$};
      \node[fill=tikzlightgray] (out1) at ($(a)!0.5!(x)+(0,-3)$) {\color{tikzred}$2*$};
      \node[fill=tikzlightgray] (out2) at ($(x)!0.5!(y)+(0,-3)$) {\color{tikzred}$3*$};
      \draw[->] (a) -- (ax);
      \draw[->] (x) -- (ax);
      \draw[->] (ax) -- (axpy);
      \draw[->] (y) -- (axpy);
      \draw[->] (axpy) -- (out1);
      \draw[->] (axpy) -- (out2);
      \node[draw=none, rectangle, anchor=south west, minimum height=0, align=left, text=tikzgreen,
          font=\tiny\linespread{0.8}\selectfont] (acounts) at ([xshift=-0.05in]a.north east) {
        \texttt{R: 2}};
      \node[draw=none, rectangle, anchor=south west, minimum height=0, align=left, text=tikzgreen,
          font=\tiny\linespread{0.8}\selectfont] (xcounts) at ([xshift=-0.05in]x.north east) {
        \texttt{R: 2}};
      \node[draw=none, rectangle, anchor=south west, minimum height=0, align=left, text=tikzgreen,
          font=\tiny\linespread{0.8}\selectfont] (ycounts) at ([xshift=-0.05in]y.north east) {
        \texttt{R: 2}};
      \node[draw=none, rectangle, anchor=west, minimum height=0, align=left, text=tikzgreen,
          font=\tiny\linespread{0.8}\selectfont] (axcounts) at (ax.east) {
        \texttt{C: 2}};
      \node[draw=none, rectangle, anchor=west, minimum height=0, align=left, text=tikzgreen,
          font=\tiny\linespread{0.8}\selectfont] (axpycounts) at (axpy.east) {
        \texttt{C: 2}};
      \node[draw=none, rectangle, anchor=north west, minimum height=0, align=left, text=tikzgreen,
          font=\tiny\linespread{0.8}\selectfont] (out1counts) at ([xshift=-0.05in]out1.south east) {
        \texttt{C: 1}\\\texttt{W: 1}};
      \node[draw=none, rectangle, anchor=north west, minimum height=0, align=left, text=tikzgreen,
          font=\tiny\linespread{0.8}\selectfont] (out2counts) at ([xshift=-0.05in]out2.south east) {
        \texttt{C: 1}\\\texttt{W: 1}};
      \node[draw=none, rectangle, minimum height=0] (totals) at ($(out1)!0.5!(out2)+(0,-1.5)$) {
        \begin{tabular}{lr}
          Reads (\texttt{{\color{tikzgreen}R}}): & 6 \\
          Writes (\texttt{{\color{tikzgreen}W}}): & 2 \\
          Computes (\texttt{{\color{tikzgreen}C}}): & 6 \\
        \end{tabular}};
      \end{tikzpicture}
    \end{minipage}
    \hfill
    \begin{minipage}{0.45\textwidth}
      \centering
      \begin{tikzpicture}[
        >=Latex,
        every node/.style={draw, circle, minimum width=0.6cm, minimum height=0.6cm},
        node distance=0.15cm and 0.15cm
      ]
      \footnotesize
      \node[fill=tikzlightgray] (a) at (-1.1,0) {$a$};
      \node[fill=tikzlightgray] (x) at (0,0) {$x$};
      \node[fill=tikzlightgray] (y) at (1.1,0) {$y$};
      \node (ax) at ($(a)!0.5!(x)+(0,-1)$) {\color{tikzred}$*$};
      \node[fill=tikzlightgray] (axpy) at ($(x)+(0,-2)$) {\color{tikzred}$+$};
      \node[fill=tikzlightgray] (out1) at ($(a)!0.5!(x)+(0,-3)$) {\color{tikzred}$2*$};
      \node[fill=tikzlightgray] (out2) at ($(x)!0.5!(y)+(0,-3)$) {\color{tikzred}$3*$};
      \draw[->] (a) -- (ax);
      \draw[->] (x) -- (ax);
      \draw[->] (ax) -- (axpy);
      \draw[->] (y) -- (axpy);
      \draw[->] (axpy) -- (out1);
      \draw[->] (axpy) -- (out2);
      \node[draw=none, rectangle, anchor=south west, minimum height=0, align=left, text=tikzgreen,
          font=\tiny\linespread{0.8}\selectfont] (acounts) at ([xshift=-0.05in]a.north east) {
        \texttt{R: 1}};
      \node[draw=none, rectangle, anchor=south west, minimum height=0, align=left, text=tikzgreen,
          font=\tiny\linespread{0.8}\selectfont] (xcounts) at ([xshift=-0.05in]x.north east) {
        \texttt{R: 1}};
      \node[draw=none, rectangle, anchor=south west, minimum height=0, align=left, text=tikzgreen,
          font=\tiny\linespread{0.8}\selectfont] (ycounts) at ([xshift=-0.05in]y.north east) {
        \texttt{R: 1}};
      \node[draw=none, rectangle, anchor=west, minimum height=0, align=left, text=tikzgreen,
          font=\tiny\linespread{0.8}\selectfont] (axcounts) at (ax.east) {
        \texttt{C: 1}};
      \node[draw=none, rectangle, anchor=west, minimum height=0, align=left, text=tikzgreen,
          font=\tiny\linespread{0.8}\selectfont] (axpycounts) at (axpy.east) {
        \texttt{C: 1}\\\texttt{W: 1}\\\texttt{R: 2}};
      \node[draw=none, rectangle, anchor=north west, minimum height=0, align=left, text=tikzgreen,
          font=\tiny\linespread{0.8}\selectfont] (out1counts) at ([xshift=-0.05in]out1.south east) {
        \texttt{C: 1}\\\texttt{W: 1}};
      \node[draw=none, rectangle, anchor=north west, minimum height=0, align=left, text=tikzgreen,
          font=\tiny\linespread{0.8}\selectfont] (out2counts) at ([xshift=-0.05in]out2.south east) {
        \texttt{C: 1}\\\texttt{W: 1}};
      \node[draw=none, rectangle, minimum height=0] (totals) at ($(out1)!0.5!(out2)+(0,-1.5)$) {
        \begin{tabular}{lr}
          Reads (\texttt{{\color{tikzgreen}R}}): & 5 \\
          Writes (\texttt{{\color{tikzgreen}W}}): & 3 \\
          Computes (\texttt{{\color{tikzgreen}C}}): & 4 \\
        \end{tabular}};
      \end{tikzpicture}
    \end{minipage}
    \caption{
      A successful materialization choice (of $ax + y$) that reduces computation without increasing memory
      accesses.
    }\label{fig:materializationsuccess}
  \end{subfigure}%
  \hfill
  \begin{subfigure}[t]{0.45\textwidth}
    \centering
    \begin{minipage}{0.45\textwidth}
      \centering
      \begin{tikzpicture}[
        >=Latex,
        every node/.style={draw, circle, minimum width=0.6cm, minimum height=0.6cm},
        node distance=0.15cm and 0.15cm
      ]
      \footnotesize
      \node[fill=tikzlightgray] (x) at (-0.6,0) {$x$};
      \node[fill=tikzlightgray] (y) at (0.6,0) {$y$};
      \node (twox) at ($(x)+(0,-1)$) {\color{tikzred}$2*$};
      \node (twoxpy) at ($(x)!0.5!(y)+(0,-2)$) {\color{tikzred}$+$};
      \node[fill=tikzlightgray] (out1) at ($(x)+(0,-3)$) {\color{tikzred}$2*$};
      \node[fill=tikzlightgray] (out2) at ($(y)+(0,-3)$) {\color{tikzred}$3*$};
      \draw[->] (x) -- (twox);
      \draw[->] (twox) -- (twoxpy);
      \draw[->] (y) -- (twoxpy);
      \draw[->] (twoxpy) -- (out1);
      \draw[->] (twoxpy) -- (out2);
      \node[draw=none, rectangle, anchor=south west, minimum height=0, align=left, text=tikzgreen,
          font=\tiny\linespread{0.8}\selectfont] (xcounts) at ([xshift=-0.05in]x.north east) {
        \texttt{R: 2}};
      \node[draw=none, rectangle, anchor=south west, minimum height=0, align=left, text=tikzgreen,
          font=\tiny\linespread{0.8}\selectfont] (ycounts) at ([xshift=-0.05in]y.north east) {
        \texttt{R: 2}};
      \node[draw=none, rectangle, anchor=south west, minimum height=0, align=left, text=tikzgreen,
          font=\tiny\linespread{0.8}\selectfont] (twoxcounts) at ([yshift=-0.05in]twox.north east) {
        \texttt{C: 2}};
      \node[draw=none, rectangle, anchor=west, minimum height=0, align=left, text=tikzgreen,
          font=\tiny\linespread{0.8}\selectfont] (twoxpycounts) at (twoxpy.east) {
        \texttt{C: 2}};
      \node[draw=none, rectangle, anchor=north west, minimum height=0, align=left, text=tikzgreen,
          font=\tiny\linespread{0.8}\selectfont] (out1counts) at ([xshift=-0.05in]out1.south east) {
        \texttt{C: 1}\\\texttt{W: 1}};
      \node[draw=none, rectangle, anchor=north west, minimum height=0, align=left, text=tikzgreen,
          font=\tiny\linespread{0.8}\selectfont] (out2counts) at ([xshift=-0.05in]out2.south east) {
        \texttt{C: 1}\\\texttt{W: 1}};
      \node[draw=none, rectangle, minimum height=0] (totals) at ($(out1)!0.5!(out2)+(0,-1.5)$) {
        \begin{tabular}{lr}
          Reads (\texttt{{\color{tikzgreen}R}}): & 4 \\
          Writes (\texttt{{\color{tikzgreen}W}}): & 2 \\
          Computes (\texttt{{\color{tikzgreen}C}}): & 6 \\
        \end{tabular}};
      \end{tikzpicture}
    \end{minipage}
    \hfill
    \begin{minipage}{0.45\textwidth}
      \centering
      \begin{tikzpicture}[
        >=Latex,
        every node/.style={draw, circle, minimum width=0.6cm, minimum height=0.6cm},
        node distance=0.15cm and 0.15cm
      ]
      \footnotesize
      \node[fill=tikzlightgray] (x) at (-0.6,0) {$x$};
      \node[fill=tikzlightgray] (y) at (0.6,0) {$y$};
      \node (twox) at ($(x)+(0,-1)$) {\color{tikzred}$2*$};
      \node[fill=tikzlightgray] (twoxpy) at ($(x)!0.5!(y)+(0,-2)$) {\color{tikzred}$+$};
      \node[fill=tikzlightgray] (out1) at ($(x)+(0,-3)$) {\color{tikzred}$2*$};
      \node[fill=tikzlightgray] (out2) at ($(y)+(0,-3)$) {\color{tikzred}$3*$};
      \draw[->] (x) -- (twox);
      \draw[->] (twox) -- (twoxpy);
      \draw[->] (y) -- (twoxpy);
      \draw[->] (twoxpy) -- (out1);
      \draw[->] (twoxpy) -- (out2);
      \node[draw=none, rectangle, anchor=south west, minimum height=0, align=left, text=tikzgreen,
          font=\tiny\linespread{0.8}\selectfont] (xcounts) at ([xshift=-0.05in]x.north east) {
        \texttt{R: 1}};
      \node[draw=none, rectangle, anchor=south west, minimum height=0, align=left, text=tikzgreen,
          font=\tiny\linespread{0.8}\selectfont] (ycounts) at ([xshift=-0.05in]y.north east) {
        \texttt{R: 1}};
      \node[draw=none, rectangle, anchor=south west, minimum height=0, align=left, text=tikzgreen,
          font=\tiny\linespread{0.8}\selectfont] (twoxcounts) at ([yshift=-0.05in]twox.north east) {
        \texttt{C: 1}};
      \node[draw=none, rectangle, anchor=west, minimum height=0, align=left, text=tikzgreen,
          font=\tiny\linespread{0.8}\selectfont] (twoxpycounts) at (twoxpy.east) {
        \texttt{C: 1}\\\texttt{W: 1}\\\texttt{R: 2}};
      \node[draw=none, rectangle, anchor=north west, minimum height=0, align=left, text=tikzgreen,
          font=\tiny\linespread{0.8}\selectfont] (out1counts) at ([xshift=-0.05in]out1.south east) {
        \texttt{C: 1}\\\texttt{W: 1}};
      \node[draw=none, rectangle, anchor=north west, minimum height=0, align=left, text=tikzgreen,
          font=\tiny\linespread{0.8}\selectfont] (out2counts) at ([xshift=-0.05in]out2.south east) {
        \texttt{C: 1}\\\texttt{W: 1}};
      \node[draw=none, rectangle, minimum height=0] (totals) at ($(out1)!0.5!(out2)+(0,-1.5)$) {
        \begin{tabular}{lr}
          Reads (\texttt{{\color{tikzgreen}R}}): & 4 \\
          Writes (\texttt{{\color{tikzgreen}W}}): & 3 \\
          Computes (\texttt{{\color{tikzgreen}C}}): & 4 \\
        \end{tabular}};
      \end{tikzpicture}
    \end{minipage}
  \caption{
    A less successful materialization choice (of $2x + y$) that reduces the amount of computation but increases
    the number of memory accesses.
  }\label{fig:materializationfail}
  \end{subfigure}%
  \Description{
    Two examples of materialization choices, each containing a `before' and `after' graph.
    The first example computes an intermediate expression from three materialized inputs, and then
    multiplies that by two different numbers to produce a pair of outputs.
    Materializing this expression keeps the memory access count the same, but reduces the amount of
    computation.
    The second examples computes an intermediate expression from two materialized inputs, and then
    multiplies that by two different numbers to produce a pair of outputs.
    Materializing this expression reduces the amount of computation, but increases the memory access
    count.
  }
  \caption{
    Two examples of materialization, showing their costs before and after in terms of array reads
    ({\color{tikzgreen}\texttt{R}}), writes ({\color{tikzgreen}\texttt{W}}), and computes
    ({\color{tikzgreen}\texttt{C}}). Unmaterialized nodes are shown in white; materialized nodes are
    shown in gray.
  }\label{fig:materialization}
\end{figure}
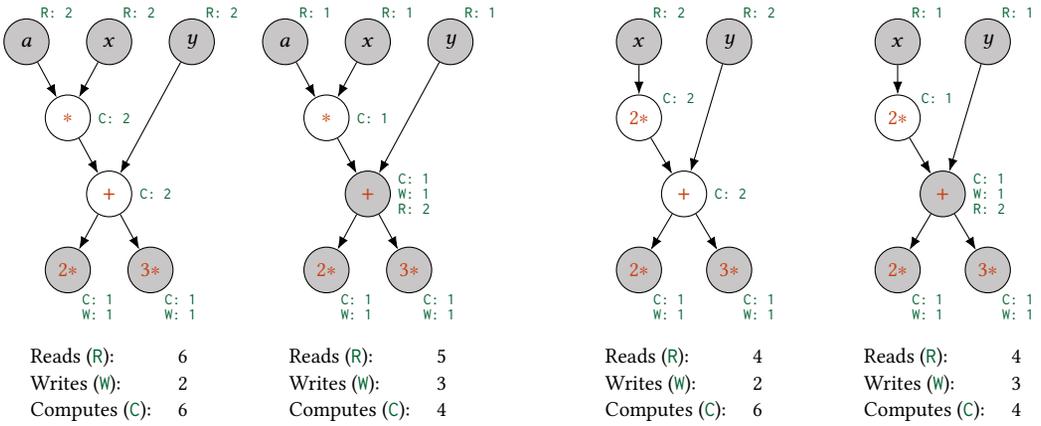

The ADFG created by the computation does not initially contain any notion of \emph{materialization},
that is, which arrays in the graph are to be stored explicitly in memory.
It is therefore the responsibility of the transformation chain to make this decision.
For the sake of arithmetic intensity, the fraction of arrays in the graph that are materialized
should remain small.
On the other hand, under-materializing can lead to excessive repeated work, as unmaterialized
arrays are evaluated every time they are referenced (see \cref{fig:materialization} for some examples).
Thus the materialization algorithm requires a balance between these two extremes.

We currently use a greedy materialization strategy based on a simple heuristic: materialize
nodes that have \emph{more than one materialized predecessor and more than one successor}.
In addition, all nodes containing reductions (in our case, typically einsum nodes)
are forcibly materialized due to their increased arithmetic intensity.

For a number of physical models, we have found this simple approach to work well: it is both straightforward to
apply and delivers acceptable results (limiting materialization rates to 5-15\% and repeated computation
to less than 10\%).
Applying this strategy to several of our more complex physical models, however, did reveal some
shortcomings:
\begin{enumerate*}[label=(\arabic*)]
\item smaller-sized arrays whose entries are used repeatedly in the evaluation of a single larger array leading to
  unnecessary recomputation due to the larger array being counted as a single successor, and
\item simple sub-expressions (fewer than two materialized predecessors) that have a large number of
  successors.
\end{enumerate*}
In at least one of the ADFGs we have considered, the missed materializations led to recomputation rates of over 75\%.
Despite this, extending the heuristic to na\"ively materialize in these cases did not prove fruitful:
computation times remained the same or increased slightly.
Given that our applications are largely memory-bandwidth-bound, it is likely that the reduction in
floating point operations from materializing in such cases is counteracted by an increase in
memory accesses. An example of how this can happen is shown in \cref{fig:materializationfail}.
We expect that future research on improved heuristics, particularly ones that
more tightly integrate with transformations applied later in the pipeline, has the
potential to lead to drastic improvements.

\subsubsection{Distributed Partitioning}

To facilitate distributed-memory computation, ADFGs containing communication nodes are
analyzed to determine a coarse-grained communication graph
in which each node represents a communication operation (both a send and a receive), and each edge
represents rank-local data flow. Individual segments are computed on each compute unit
and then sent to a coordinating compute unit for processing.
This coordinator computes a topological ordering of communication operations and greedily groups them
into batches (\cref{fig:distpart1}).
After the communication batches are formed, they are distributed back out to all
compute units, where they are subsequently treated as boundaries between the
to-be-constructed node-local computational sub-ADFGs.
Computational nodes from the full ADFG are inserted between batches:
\begin{enumerate*}[label=(\arabic*)]
\item after all of their prerequisite receives have been completed, and
\item just before their outputs are needed (\cref{fig:distpart2}).
\end{enumerate*}
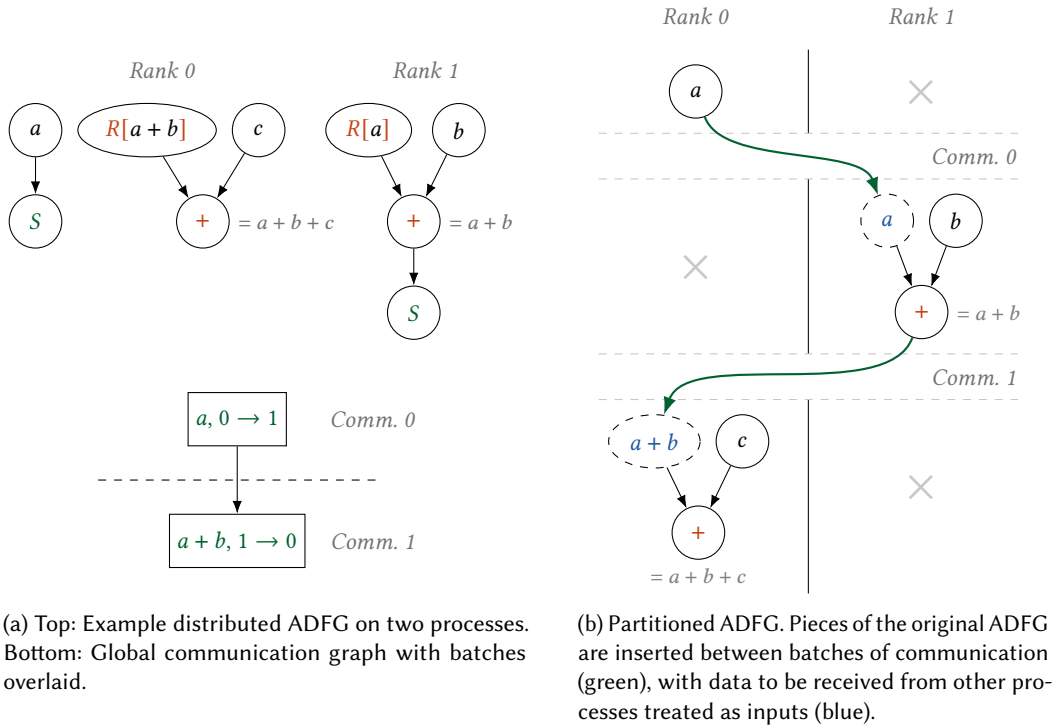
\begin{figure}[!thp]
  \centering
  \begin{subfigure}[t]{0.50\textwidth}
    \centering
    \begin{tikzpicture}[
      >=Latex,
      every node/.style={draw, circle, minimum width=0.7cm, minimum height=0.7cm},
      node distance=0.2cm and 0.2cm
    ]

    \small
    \node[draw=none] (adfganchor) at (0,2.7) {};
    \node[draw=none] (rank0anchor) at ($(adfganchor)+(-1,0)$) {};
    \node[draw=none] (rank1anchor) at ($(adfganchor)+(2.7,0)$) {};
    \node[draw=none] (rank0label) at (rank0anchor) {\color{tikzgray}\textit{Rank 0}};
    \node[ellipse] (recvaplusb) at ($(rank0anchor)+(-0.2,-0.8)$) {\color{tikzred}$R[{\color{black}a+b}]$};
    \node[right=of recvaplusb] (c) {$c$};
    \node[left=of recvaplusb] (a) {$a$};
    \node (senda) at ($(a)+(0,-1.2)$) {\color{tikzgreen}$S$};
    \node[ellipse] (aplusbplusc) at ($(recvaplusb)!0.5!(c)+(0,-1.2)$) {\color{tikzred}$+$};
    \node[draw=none, rectangle, anchor=west, minimum height=0] (aplusbplusclabel) at (aplusbplusc.east) {
      \color{tikzgray}\footnotesize$=a+b+c$};
    \draw[->] (a) -- (senda);
    \draw[->] (recvaplusb) -- (aplusbplusc);
    \draw[->] (c) -- (aplusbplusc);
    \node[draw=none] (rank1label) at ($(rank1anchor)+(-0.2,0)$) {\color{tikzgray}\textit{Rank 1}};
    \node[ellipse] (recva) at ($(rank1anchor)+(-0.95,-0.8)$) {\color{tikzred}$R[{\color{black}a}]$};
    \node (b) at ($(recva)+(1.2,0)$) {$b$};
    \node[ellipse] (aplusb) at ($(recva)!0.5!(b)+(0,-1.2)$) {\color{tikzred}$+$};
    \node[draw=none, rectangle, anchor=west, minimum height=0] (aplusblabel) at (aplusb.east) {
      \color{tikzgray}\footnotesize$=a+b$};
    \node (sendaplusb) at ($(aplusb)+(0,-1.2)$) {\color{tikzgreen}$S$};
    \draw[->] (recva) -- (aplusb);
    \draw[->] (b) -- (aplusb);
    \draw[->] (aplusb) -- (sendaplusb);

    \small
    \node[draw=none] (commbatchanchor) at (0,-2.7) {};
    \node[rectangle] (comm1) at ($(commbatchanchor)+(0,0.8)$) {\color{tikzgreen}$a$, $0\rightarrow 1$};
    \node[rectangle] (comm2) at ($(commbatchanchor)+(0,-0.8)$) {\color{tikzgreen}$a+b$, $1\rightarrow 0$};
    \node[draw=none] (comm0label) at ($(commbatchanchor)+(1.8,0.8)$) {\color{tikzgray}\textit{Comm. 0}};
    \node[draw=none] (comm1label) at ($(commbatchanchor)+(1.8,-0.8)$) {\color{tikzgray}\textit{Comm. 1}};
    \draw[dashed] ($(commbatchanchor)+(-1.85,0)$) -- ($(commbatchanchor)+(1.85,0)$);
    \draw[->] (comm1) -- (comm2);

    \end{tikzpicture}
    \caption{
      Top: Example distributed ADFG on two processes.
      Bottom: Global communication graph with batches overlaid.
  }\label{fig:distpart1}
  \end{subfigure}%
  \hfill
  \begin{subfigure}[t]{0.45\textwidth}
    \centering
    \begin{tikzpicture}[
      >=Latex,
      every node/.style={draw, circle, minimum width=0.7cm, minimum height=0.7cm},
      node distance=0.2cm and 0.2cm
    ]
    \small
    \node[draw=none] (rank0label) at (-1.5,0) {\color{tikzgray}\textit{Rank 0}};
    \node[draw=none] (rank1label) at (1.5,0) {\color{tikzgray}\textit{Rank 1}};
    \node[draw=none] (part0anchor) at (0,-1) {};
    \node[draw=none] (rank0part0anchor) at ($(part0anchor)+(-1.5,0)$) {};
    \node[draw=none] (rank1part0anchor) at ($(part0anchor)+(1.5,0)$) {};
    \node (a) at (rank0part0anchor) {$a$};
    \node[draw=none] (rank1part0) at (rank1part0anchor) {\huge\color{tikzlightgray}$\times$};
    \node[draw=none] (part1anchor) at (0,-3.3) {};
    \node[draw=none] (rank0part1anchor) at ($(part1anchor)+(-1.5,0)$) {};
    \node[draw=none] (rank1part1anchor) at ($(part1anchor)+(1.5,0)$) {};
    \node[draw=none] (rank0part1) at (rank0part1anchor) {\huge\color{tikzlightgray}$\times$};
    \node[dashed, ellipse] (inputa) at ($(rank1part1anchor)+(-0.45,0.6)$) {\color{tikzblue}$a$};
    \node[right=of inputa] (b) {$b$};
    \node (aplusb) at ($(inputa)!0.5!(b)+(0,-1.2)$) {\color{tikzred}$+$};
    \node[draw=none, rectangle, anchor=west, minimum height=0] (aplusblabel) at (aplusb.east) {
      \color{gray}\footnotesize$=a+b$};
    \draw[->] (inputa) -- (aplusb);
    \draw[->] (b) -- (aplusb);
    \node[draw=none] (part2anchor) at (0,-6.2) {};
    \node[draw=none] (rank0part2anchor) at ($(part2anchor)+(-1.5,0)$) {};
    \node[draw=none] (rank1part2anchor) at ($(part2anchor)+(1.5,0)$) {};
    \node[dashed, ellipse] (inputaplusb) at ($(rank0part2anchor)+(-0.55,0.6)$) {\color{tikzblue}$a+b$};
    \node[right=of inputaplusb] (c) {$c$};
    \node (aplusbplusc) at ($(inputaplusb)!0.5!(c)+(0,-1.2)$) {\color{tikzred}$+$};
    \node[draw=none, rectangle, anchor=north, minimum height=0] (aplusbplusclabel) at (aplusbplusc.south) {
      \color{gray}\footnotesize$=a+b+c$};
    \draw[->] (inputaplusb) -- (aplusbplusc);
    \draw[->] (c) -- (aplusbplusc);
    \node[draw=none] (rank1part2) at (rank1part2anchor) {\huge\color{tikzlightgray}$\times$};
    \draw[-] ($(part0anchor)+(0,0.55)$) -- ($(part0anchor)+(0,-0.55)$);
    \draw[-] ($(part1anchor)+(0,1.15)$) -- ($(part1anchor)+(0,-1.15)$);
    \draw[-] ($(part2anchor)+(0,1.15)$) -- ($(part2anchor)+(0,-1.4)$);
    \draw[dashed, color=tikzlightgray] ($(part0anchor)+(-2.79,-0.55)$) -- ($(part0anchor)+(2.79,-0.55)$);
    \draw[dashed, color=tikzlightgray] ($(part1anchor)+(-2.79,1.15)$) -- ($(part1anchor)+(2.79,1.15)$);
    \draw[dashed, color=tikzlightgray] ($(part1anchor)+(-2.79,-1.15)$) -- ($(part1anchor)+(2.79,-1.15)$);
    \draw[dashed, color=tikzlightgray] ($(part2anchor)+(-2.79,1.15)$) -- ($(part2anchor)+(2.79,1.15)$);
    \draw[->,thick,color=tikzgreen] (a) to[out=-70,in=110,out looseness=0.7,in looseness=0.7] (inputa);
    \draw[->,thick,color=tikzgreen] (aplusb) to[out=-110,in=70,out looseness=0.7,in looseness=0.7] (inputaplusb);
    \node[draw=none] (comm0label) at (2.2,-1.85) {\color{tikzgray}\textit{Comm. 0}};
    \node[draw=none] (comm1label) at (2.2,-4.75) {\color{tikzgray}\textit{Comm. 1}};
    \end{tikzpicture}
    \caption{
      Partitioned ADFG. Pieces of the original ADFG are inserted between batches of communication
      (green), with data to be received from other processes treated as inputs (blue).}\label{fig:distpart2}
  \end{subfigure}%
  \Description{The left portion of the figure has two two-level binary trees representing an addition operation on ranks 0 and 1.  Rank 0 has the operation of $a+b$ being added to $c$; rank 1 shows the operation of computing $a+b$.  The right portion of the figure shows the coordination of the input/output of $a$, $a+b$, and $a+b+c$, with $a$ residing on rank 0 being input to $a+b$ on rank 1, followed by the output of $a+b$ as input to $a+b+c$ on rank 0.}
  \caption{
    Partitioning a distributed ADFG. Note: For this two-rank example, destination ranks of sends
      ({\color{tikzgreen}$S$}) and source ranks of receives ({\color{tikzred}$R[{\color{black}\ldots}]$})
      are trivial and thus omitted.
  }\label{fig:distpart}
\end{figure}

Constructing the communication graph globally guarantees that the partitioning process will not
result in any circular dependencies between node-local per-batch compute-only ADFGs.
A further benefit of the global communication graph is that it enables analyses to check for mismatched communication.
While this solution is certainly not asymptotically scalable, in practice, we have
not found this choice to result in any significant runtime burden at problem scales
of relevance.

\subsubsection{Concatenation of Function Calls}\label{sec:concatenation}

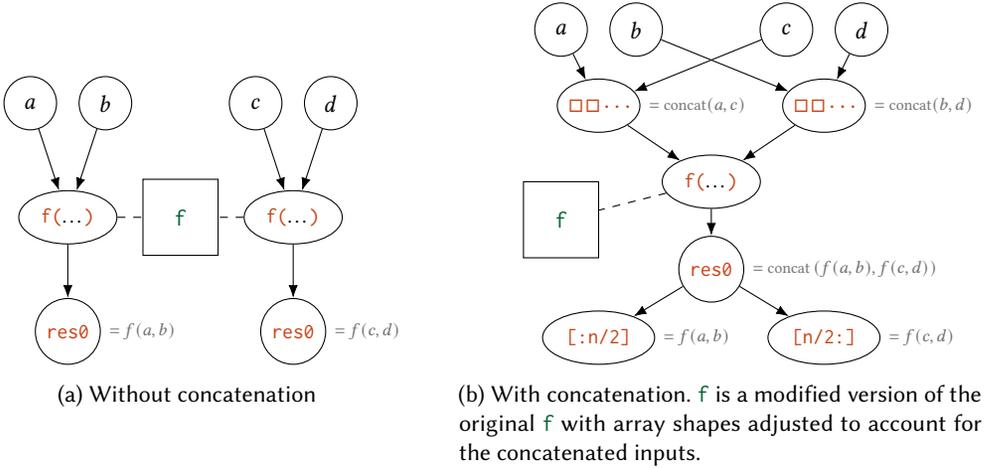
\begin{figure}[!thp]
  \centering
  \begin{subfigure}[t]{0.48\textwidth}
    \centering
    \begin{tikzpicture}[
      >=Latex,
      every node/.style={draw, circle, minimum width=0.7cm, minimum height=0.7cm},
      node distance=0.2cm and 0.2cm
    ]
    \small
    \node[draw=none] (whitespace) at (-2.5,0) {};
    \node (a) at (-2,-0.6) {$a$};
    \node (b) at (-1, -0.6) {$b$};
    \node (c) at (1, -0.6) {$c$};
    \node (d) at (2, -0.6) {$d$};
    \node[ellipse] (callf1) at ($(a)!0.5!(b)+(0,-1.5)$) {\footnotesize\color{tikzred}\code{f(}{\color{black}\ldots}\code{)}};
    \node (result1) at ($(callf1)+(0,-1.5)$) {\footnotesize\color{tikzred}\code{res0}};
    \node[draw=none, rectangle, anchor=west, minimum height=0] (result1label) at (result1.east) {
      \color{tikzgray}\tiny$=f(a, b)$};
    \node[ellipse] (callf2) at ($(c)!0.5!(d)+(0,-1.5)$) {\footnotesize\color{tikzred}\code{f(}{\color{black}\ldots}\code{)}};
    \node (result2) at ($(callf2)+(0,-1.5)$) {\footnotesize\color{tikzred}\code{res0}};
    \node[draw=none, rectangle, anchor=west, minimum height=0] (result2label) at (result2.east) {
      \color{tikzgray}\tiny$=f(c, d)$};
    \node[rectangle, minimum width=1cm, minimum height=1cm] (f) at (0, -2.1) {\color{tikzgreen}\code{f}};
    \draw[->] (a) -- (callf1);
    \draw[->] (b) -- (callf1);
    \draw[->] (c) -- (callf2);
    \draw[->] (d) -- (callf2);
    \draw[->] (callf1) -- (result1);
    \draw[->] (callf2) -- (result2);
    \draw[dashed] (callf1) -- (f);
    \draw[dashed] (callf2) -- (f);
    \end{tikzpicture}
    \caption{Without concatenation}
  \end{subfigure}
\hfill
  \begin{subfigure}[t]{0.5\textwidth}
    \centering
    \begin{tikzpicture}[
      >=Latex,
      every node/.style={draw, circle, minimum width=0.7cm, minimum height=0.7cm},
      node distance=0.2cm and 0.2cm
    ]
    \small
    \node[draw=none] (whitespace) at (-3,0) {};
    \node (a) at (-2,-0.6) {$a$};
    \node (b) at (-1, -0.6) {$b$};
    \node (c) at (1, -0.6) {$c$};
    \node (d) at (2, -0.6) {$d$};
    \node[ellipse, minimum width=1.1cm] (ac) at ($(a)!0.5!(b)+(0,-1)$) {};
    \node[draw=none] at ($(ac.center) + (-0.3,0)$) {\color{tikzred}$\square$};
    \node[draw=none] at ($(ac.center) + (-0.075,0)$) {\color{tikzred}$\square$};
    \node[draw=none] at ($(ac.center) + (0.275,-0.05)$) {\color{tikzred}$\cdots$};
    \node[draw=none, rectangle, anchor=west, minimum height=0] (aclabel) at (ac.east) {
      \color{tikzgray}\tiny$=\text{concat}(a,c)$};
    \node[ellipse, minimum width=1.1cm] (bd) at ($(c)!0.5!(d)+(0,-1)$) {};
    \node[draw=none] at ($(bd.center) + (-0.3,0)$) {\color{tikzred}$\square$};
    \node[draw=none] at ($(bd.center) + (-0.075,0)$) {\color{tikzred}$\square$};
    \node[draw=none] at ($(bd.center) + (0.275,-0.05)$) {\color{tikzred}$\cdots$};
    \node[draw=none, rectangle, anchor=west, minimum height=0] (bdlabel) at (bd.east) {
      \color{tikzgray}\tiny$=\text{concat}(b,d)$};
    \node[ellipse] (callf) at ($(ac)!0.5!(bd)+(0,-1)$) {\footnotesize\color{tikzred}\code{f(}{\color{black}\ldots}\code{)}};
    \node (result) at ($(callf)+(0,-1.15)$) {\footnotesize\color{tikzred}\code{res0}};
    \node[draw=none, rectangle, anchor=west, minimum height=0] (resultlabel) at (result.east) {
      \color{tikzgray}\tiny$=\text{concat}\left(f(a, b),f(c, d)\right)$};
    \node[ellipse] (split1) at ($(result)+(-1.5,-0.9)$) {\footnotesize\color{tikzred}\code{[:n/2]}};
    \node[draw=none, rectangle, anchor=west, minimum height=0] (split1label) at (split1.east) {
      \color{tikzgray}\tiny$=f(a, b)$};
    \node[ellipse] (split2) at ($(result)+(1.5,-0.9)$) {\footnotesize\color{tikzred}\code{[n/2:]}};
    \node[draw=none, rectangle, anchor=west, minimum height=0] (split2label) at (split2.east) {
      \color{tikzgray}\tiny$=f(c, d)$};
    \node[rectangle, minimum width=1cm, minimum height=1cm] (f) at (-2, -3.1) {\color{tikzgreen}\code{f}};
    \draw[->] (a) -- (ac);
    \draw[->] (b) -- (bd);
    \draw[->] (c) -- (ac);
    \draw[->] (d) -- (bd);
    \draw[->] (ac) -- (callf);
    \draw[->] (bd) -- (callf);
    \draw[->] (callf) -- (result);
    \draw[->] (result) -- (split1);
    \draw[->] (result) -- (split2);
    \draw[dashed] (callf) -- (f);
    \end{tikzpicture}
    \caption{
      With concatenation. {\color{tikzgreen}\code{f}} is a modified version of the original
      {\color{tikzgreen}\code{f}} with array shapes adjusted to account for the concatenated inputs.}
  \end{subfigure}
  \Description{On the left are two tree, representing $f(a,b)$ and $f(c,d)$.  On the right is a combined tree representing concatenated inputs $(a,c)$ and $(b,d)$ to the function calls.}
  \caption{Transforming an ADFG by concatenation of function calls. For simplicity, this example
    assumes $a$, $b$, $c$, and $d$ are one-dimensional arrays of equal length $n$.
    }
  \label{fig:concat}
\end{figure}

If the operations inside an outlined function (see \cref{sec:outlining}) are sufficiently
`well-behaved' (see below), it is possible to transform a collection of independent calls into a single call by
concatenating the inputs, outputs, and function body nodes of each call along suitable axes to
create a single, larger function call (as shown in~\cref{fig:concat}).
This transformation lowers computational cost by reducing the number of kernel invocations and increasing
the size of the arrays being operated on, improving GPU utilization.

The definition of `well-behaved' depends on a number of specifics.
At the outset, all calls to specific user-identified functions are considered
eligible for concatenation. Concatenation groups are then split to ensure that
all concatenated invocations are independent. Finally, an axis matching procedure
determines axes along which concatenation can take place while leaving the remaining
array dimensions unchanged. We have explored a number of different algorithms for
this matching, varying by their tolerance to varying axis positions. Ultimately,
a relatively simple algorithm based on connecting axes at the input of a node
to those at the output proved sufficient.

\subsection{Stage 3: From Array-Level IR to Scalar IR}

After the ADFG has been transformed and optimized, the next stage rewrites the
graph into a lower level intermediate representation, typically expressed
in the \software{Loopy} programming model (see~\parencite{loopy} for details). This stage introduces loop
structures, temporaries, and control flow, and makes execution order and memory
usage explicit for the first time. While the ADFG deals with whole-array
expressions and symbolic graphs, the \software{Loopy} model operates on a scalar,
polyhedral model of computation, where computations are indexed over loop
domains, memory layout is defined precisely, and
execution order can be controlled via dependencies and tags~---~see~\cref{fig:code-example-loopy} for an example.
This lowering stage operates by converting each node to \emph{array comprehension}
form, in which a formula for each array entry is given in terms of entries of
its materialized predecessors. Once these array comprehensions are available,
they are straightforwardly turned into assignment statements, each
wrapped in a separate loop nest.
\begin{figure}[!thp]
\begin{lstlisting}[language=Python, xleftmargin=0.2\textwidth, xrightmargin=0.2\textwidth]
knl = lp.make_kernel(
    "{ [i,j,k]: 0 <= i < 3 and 0 <= j < 3 and 0 <= k < n_dot }",
    """
    dot = sum(k, w[k] * x[k])
    result[i * 3 + j] = x[idx[i,j]] + dot
    """
)
\end{lstlisting}
\Description{Code listing an operation with loop bounds.}
  \caption{An example \software{Loopy} kernel.}\label{fig:code-example-loopy}
\end{figure}

\subsection{Stage 4: Scalar IR Transformations}

With the low-level IR in place, additional transformations are applied
to target parallelization opportunities and optimizations in the memory
hierarchy.  In particular, \textit{loop fusion} is critical to reduce the
potentially large number of separate loops generated in the preceding stage.

In order to perform loop fusion, we rely on descriptive metadata that is
(sparsely) attached to array axes in the ADFG. In our case, axis metadata is
attached within the discontinuous finite element infrastructure, not within
the application-level code. The main purpose of this metadata is to identify
types of array axes, for example, whether an axis is indexed by the element number
of a certain discretization, the physical dimension, or the
degree of freedom inside an element of a certain local element type. This metadata is then propagated to all
`connected' axes in the ADFG and survives into the scalar IR as metadata
attached to loops.

Loops with matching metadata become \emph{fusion candidates}. Across each set of
candidates, a greedy loop fusion algorithm from the literature \cite{kennedy_1994_loopfusion},
with minor modifications, selects a suitable set, while considering the nature
of the dependencies among loop nests.

Next, we perform \emph{array contraction}. As an example, consider the loop nest
in ~\cref{fig:array-contraction-example}.
Since the outbound dependencies surrounding the variable \texttt{tmp} are elementwise,
only a single entry of the array is needed within each trip through the loop,
allowing the array reference \texttt{tmp[i]} to be replaced with a scalar temporary
\texttt{tmp}. We again follow a simple approach
from the literature~\cite{lam_1997_contractions} to realize this transformation.

It is useful to recgonize that this step will effectively undo some of the
materialization decisions made earlier in the pipeline (cf.~\cref{sec:materialization}),
once the profitability of the corresponding opportunities has become apparent
through, chiefly, loop fusion. In future work, we hope to pursue a more integrated
approach that will very likely help to reduce transform cost.
\begin{figure}[!thp]
\begin{subfigure}[b]{0.45\textwidth}
\begin{lstlisting}[language=Python, xleftmargin=0.2\textwidth, xrightmargin=0.2\textwidth]
tmp = [None] * n
z = [None] * n
for i in range(n):
  tmp[i] = x[i] + y[i]
  z[i] = 2*tmp[i]
\end{lstlisting}
\caption{Before contraction.}
\end{subfigure}
\hfill
\begin{subfigure}[b]{0.45\textwidth}
\begin{lstlisting}[language=Python, xleftmargin=0.2\textwidth, xrightmargin=0.2\textwidth]
z = [None] * n
for i in range(n):
  tmp = x[i] + y[i]
  z[i] = 2*tmp
\end{lstlisting}
\caption{After contraction.}
\end{subfigure}
\Description{Two code blocks with \code{tmp} originally represented as an array and then \code{tmp} as a scalar after contraction.}
\caption{An illustration of array contraction using Python pseudocode. Contraction replaces the intermediate array \texttt{tmp} with a scalar temporary, reducing memory usage.}
\label{fig:array-contraction-example}
\end{figure}

A final use of the propagated array axis/loop variable metadata allows us to identify
candidates for parallelization, mostly via mapping to the hardware-parallel axes
available in the GPU programming model.

\subsection{Stage 5: Emit Target Code / OpenCL}

After the transformations described in the previous section, the code in the
scalar IR has arrived at the same level of abstraction as conventional OpenCL/CUDA
code. In our case, OpenCL code is generated and just-in-time compiled via
various vendor implementations, particularly \software{PoCL}~\cite{jaaskelainen_pocl_2014}
and the RoCM AMD GPU OpenCL ICD. \Cref{fig:code-example-all} shows an
example of the type of code generated in a simple setting.

This generation is mostly straightforward,
although a number of aspects account for about 25\% in \cref{fig:compile-time-chart}.
Contributing factors include:
\begin{enumerate*}[label=(\arabic*)]
\item The amount of code generated. A typical workload as described
  in the subsequent section will emit about 100 GPU kernels, across about
  10,000 lines of code.
\item It is worth noting that \software{loopy}'s IR is itself unordered,
  necessitating an ordering pass with associated validity checking. (This
  appears in the figure as `linearization'.)
\item Since loopy is based on the polyhedral model, many operations involve
  core polyhedral primitives. While fairly low-cost individually, their sheer
  number accounts for considerable cost.
\end{enumerate*}

For NVIDIA devices, we typically eschew NVIDIA's vendor implementation of OpenCL,
preferring \software{PoCL}'s CUDA target. This helps us avoid various observed
performance limitations as well as the notable absence of a profiling capability
on the vendor implementation. \software{PoCL}'s NVIDIA target makes use of LLVM's NVPTX backend,
which generates NVIDIA's proprietary PTX intermediate language. The latter is then
just-in-time compiled into a device binary for use on the hardware via the GPU driver.
To help the viability of this approach, we have contributed a number of developments
to \software{PoCL}'s NVIDIA target~\cite{pocl_16_release_notes}, notably around the use of
32-bit pointer arithmetic for scratchpad/shared memory, the use of unified/virtual memory, as well as the resolution of
a pointer aliasing issue relating to scratchpad/shared memory.

\subsection{Stage 6: Execute}

Final code execution in the case of a single compute unit is straightforward
and occurs by enqueuing OpenCL kernel
invocations from optimized, generated Python code (an `invoker' in \software{loopy}).
Our use of OpenCL allowed the use of a large range of compute hardware through a
unified runtime, though with a wide array of actual vendor implementations,
including \software{PoCL} (CPU and NVIDIA GPU),  AMD ROCm, NVIDIA CL, Intel CL, and Apple CL.

\subsubsection{Distributed Execution}

To execute a partitioned computation graph, the execution engine first
posts non-blocking (MPI) receive operations for all expected incoming data across
all communication batches.
Then, it proceeds through an event-driven loop in which any ADFG part local
to the compute unit is eagerly executed as soon as its depenencies (local and remote)
are satisfied. Once the GPU completes computation on the ADFG part, in turn,
any necessary send operations are posted, again in a non-blocking, MPI-based
fashion. This process continues until execution of all ADFG parts is complete.
Resource management is handled through reference counting of
arrays, automatically reclaiming memory when arrays are no longer needed
by any remaining parts.

Opportunities remain for additional efficiencies, namely in overlapping
computation and communication.  For example, sends are not performed as soon as
the data is ready, only after the current part ends.  Likewise, some
careful attention to scheduling could likely eliminate some measure of GPU idle time.

\subsubsection{Kernel Profiling Support}

Kernel profiling in \mirge{} is enabled through the built-in event
profiling capabilities in OpenCL, including functionality to track and measure the
execution time of kernels during computation using specialized profiling array contexts.
Profiling is managed by storing OpenCL events related to the start and
stop times of kernel executions. At present, this accounting is performed
at the granularity of a GPU kernel launch. Further research is needed
in order to aid the user in mapping cost to individual segments of the ADFG.

\section{Applications}\label{sec:applications}

The \mirge{} approach is internally complex, but this is balanced with an outward focus on
productivity: our intent is to deliver optimized multiphysics simulations on a range of
architectures without burdening application developers with hardware idiosyncrasies.
In this section we provide numerical examples that highlight the use of
\mirge{} in practice. We demonstrate it through \mirgecom{}~\cite{mirgecom},
the simulation library for the Center for Exascale-enabled Scramjet Design
(CEESD)~\cite{ceesd} at the University of Illinois Urbana--Champaign.
\mirgecom{} is a portable, massively parallel DG library capable of simulating
reactive mixture viscous flows and coupled material responses, with a main
objective of predicting the internal behavior of scramjet combustors.

We consider three main cases:
  \begin{enumerate}
    \item an isentropic vortex in an inviscid flow,
    \item flows exhibiting shock and boundary-layers, and
    \item a large-scale scramjet simulation.
  \end{enumerate}

\subsection{\mirgecom{}}

The stack of packages that implements the stages of the
previous sections in \mirgecom{} includes \software{pytato}\footnote{\url{https://github.com/inducer/pytato}} and \software{arraycontext}\footnote{\url{https://github.com/inducer/arraycontext}}, which facilitate the description of the code as an ADFG, and
\software{loopy}\footnote{\url{https://github.com/inducer/loopy}}, which  provides the IR and transformation tools. The
mesh description and DG components are implemented with the help of
\software{meshmode}\footnote{\url{https://github.com/inducer/meshmode}}, \software{modepy}\footnote{\url{https://github.com/inducer/modepy}}, and \software{grudge}\footnote{\url{https://github.com/inducer/grudge}}.

For distributions across multiple processing
elements, \mirgecom{} uses host-based MPI through \software{mpi4py} Python
bindings. Portability is achieved with a lightweight installation utility,
\software{emirge}~\cite{emirge} which installs a Conda environment containing
components that implement \mirge{}, along with the \mirgecom{} interface. The
resulting environment supports out-of-the-box execution on most systems; more details can be found at \url{https://github.com/illinois-ceesd/mirgecom}.

\subsection{\mirgecom{} vs. \software{MFEM}}

In these experiments, we compare the runtime performance of \mirgecom{} to that of the open-source finite-element library \software{MFEM}~\cite{mfem}. We present timing data for both codes, yet it is important to recognize that strictly maximizing performance is not the only goal.   This is only a general point of comparison since the two applications differ significantly in their design, implementation, user experience, and portability.   \software{MFEM} is a \CPP{} library with tuned optimizations for both CPU and GPU platforms.  It is a reference for broadly assessing the efficiency of \mirgecom{} relative to an established package.  Analysis of the performance of the transformed code at the core of \mirge{} is reported elsewhere (for example, see~\parencite{loopy}).

We first assess the baseline CPU performance of \mirgecom{} on a canonical flow benchmark case: an isentropic vortex in an inviscid flow on a periodic domain.  This flow application is a packaged \software{MFEM} example.  In both implementations, the Euler equations for compressible inviscid flow are solved using a DG discretization using tensor product elements (TPEs), and advanced in time with explicit 4th-order Runge--Kutta. GPU execution is not supported directly in \software{MFEM}'s inviscid vortex example, so this comparison is presented on an Apple M2 processor.

\begin{figure}[!thp]
  \centering
  \includegraphics{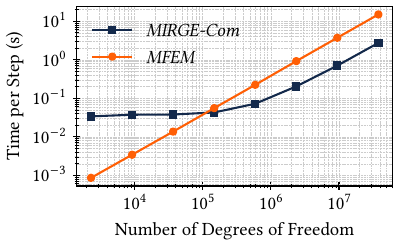}
  \hfill
  \includegraphics{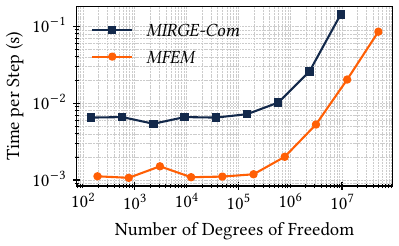}
  \Description{A log-log scale figure showing near linear growth (for larger number of DOFs) in time per time-step for MFEM and MIRGE-Com on CPU.  A log-log scale figure showing near linear growth (for large number of DOFs) in time per time-step MFEM and MIRGE-Com on GPU.}
  \caption{Mesh scaling ($h$-refinement) for \software{MFEM} and \mirgecom{}. Left: inviscid vortex on an Apple M2 processor for $p=4$ TPEs, and fixed step 4th-order Runge--Kutta (RK4). Right: scalar advection equation using $p=1$ TPEs on an AMD MI300A.}\label{fig:mirgevsmfem}
\end{figure}

We note that it is also straightforward to run \mirgecom{} on GPUs, which is a more attractive target platform for modern applications. To demonstrate GPU execution, we compare the runtime performance of \mirgecom{} to that of \software{MFEM} for the scalar advection equation in a periodic domain. GPU execution is supported in \software{MFEM} for this example application through the use of a partial assembly of the DG operators, greatly reducing the memory footprint at the cost of added computation.  \mirgecom{}, on the other hand, does full assembly of all DG operators, at the expense of increased memory footprint and traffic.
The time per time-step for both \mirgecom{} and \software{MFEM} are shown in~\cref{fig:mirgevsmfem} for both CPU and GPU execution.  For smaller numbers of elements, MIRGE has a clear performance floor, which is associated with the cost of the OpenCL kernel dispatch layer in \mirgecom{}, which is used on both CPU and GPU platforms.  A GPU performance floor is also apparent in the \software{MFEM} GPU execution time.  \software{MFEM} uses AMD's Heterogeneous-computing Interface for Portability (HIP) API~\cite{hip} to implement GPU-ready kernels, which are statically compiled.

\subsection{Shock Cases vs. \software{Ryujin}}

We next consider cases with shocks, where overintegration, artificial
viscosity, and shock limiting are needed to stabilize the flow. These more
complex flows and discretizations lead to correspondingly complex ADFGs to
process, which help test the robustness of the \mirge{} approach.

As a reference point, we implement experiments in both \mirgecom{} and in
the~\software{Ryujin} package~\cite{ryujin-1,ryujin-2}, which is an application of
\software{deal.ii}~\cite{dealii} for compressible flow.

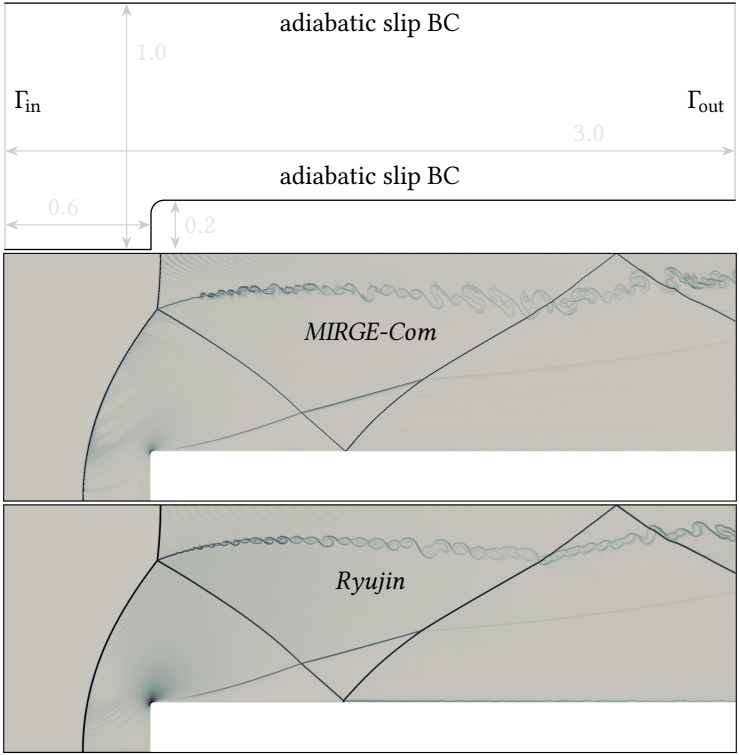
\begin{figure}[!thp]
\begin{tikzpicture}[x=92.16667pt,y=92.16667pt]   %
  \begin{scope}[shift={(0pt, 0pt)}]
    \draw[draw=black!20, line width=0.5] (0,0) -- (0, 1);
    \draw[draw=black!20, line width=0.5] (3,0.2) -- (3, 1);
    \draw[draw=black,line width=0.5,line cap=rect] (0,0) -- (0.6, 0) {[rounded corners=5pt] -- (0.6,0.2)} -- (3,0.2);
    \draw[draw=black,line width=0.5,line cap=rect] (3,1) -- (0,1);
    \node[anchor=north] at (1.5,1.0) {adiabatic slip BC};
    \node[anchor=south] at (1.5,0.2) {adiabatic slip BC};
    \node[anchor=west] at (0,0.6) {$\Gamma_{\textnormal{in}}$};
    \node[anchor=east] at (3,0.6) {$\Gamma_{\textnormal{out}}$};

    \draw[{Stealth[]}-{Stealth[]},black!20,shorten <=0.25pt,shorten >=0.25pt] (0,0.4) -- (3, 0.4) node[pos=0.8,above,black!10] {$3.0$};
    \draw[{Stealth[]}-{Stealth[]},black!20,shorten <=0.25pt,shorten >=0.25pt] (0,0.1) -- (0.6, 0.1) node[pos=0.4,above,black!10] {$0.6$};
    \draw[{Stealth[]}-{Stealth[]},black!20,shorten <=0.25pt,shorten >=0.25pt] (0.5,0) -- (0.5, 1.0) node[pos=0.8,right,black!10] {$1.0$};
    \draw[{Stealth[]}-{Stealth[]},black!20,shorten <=0.25pt,shorten >=0.25pt] (0.7,0) -- (0.7, 0.2) node[pos=0.5,right,black!10] {$0.2$};
  \end{scope}
\end{tikzpicture}

      \begin{tikzpicture}
        \node[inner sep=0pt] (image) at (0,0) {
          \includegraphics[width=0.7\textwidth,frame]{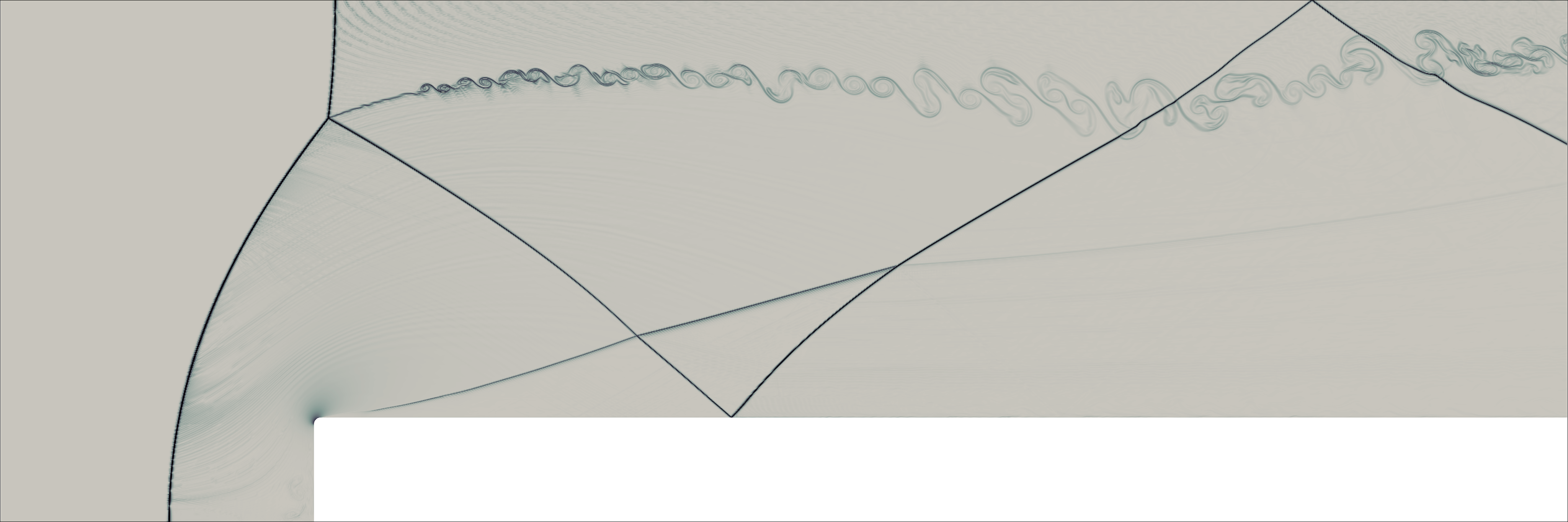}
        };
        \node at (0,0.6) {\mirgecom{}};
      \end{tikzpicture}
      \begin{tikzpicture}
        \node[inner sep=0pt] (image) at (0,0) {
          \includegraphics[width=0.7\textwidth,frame]{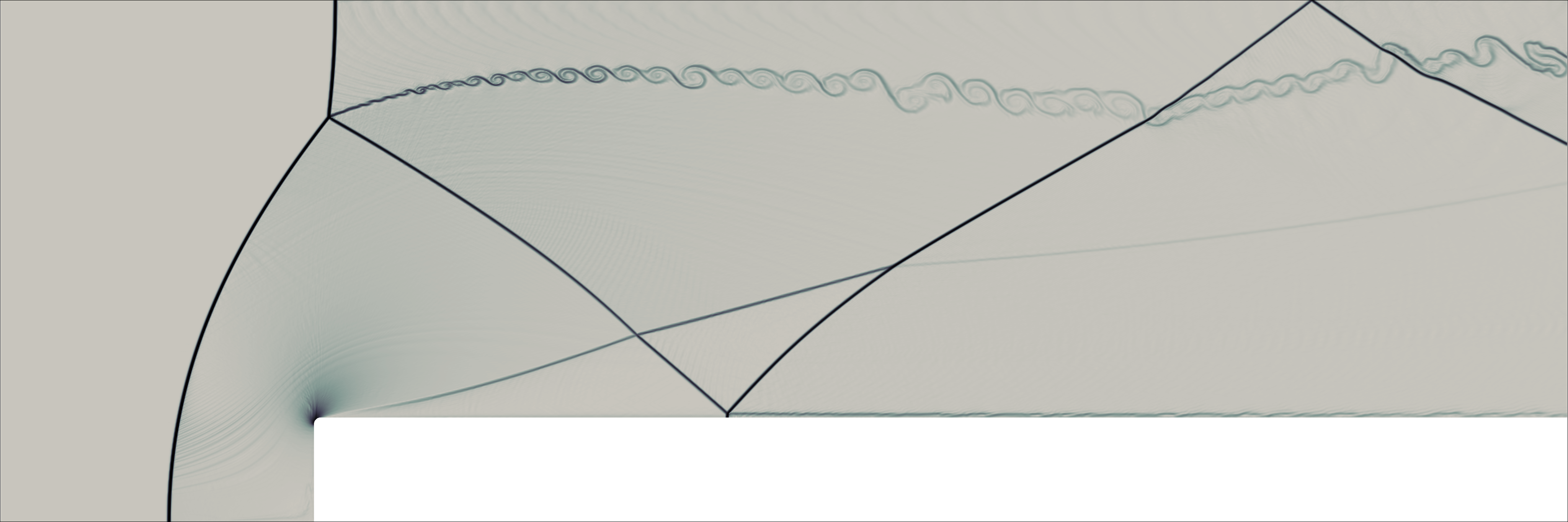}
        };
        \node at (0,0.6) {\software{Ryujin}};
      \end{tikzpicture}
  \Description{Schematic of the domain with a forward facing step.  The figure shows the dimensions to be 3 by 1, with a step at 0.6 at 0.2 high. Two figures, one for MIRGE-Com and one for Ryujin, show similar profiles of a developed flow, with shocks and small fluid features.}
  \caption{Inviscid flow past a forward-facing step. Problem setup is shown at top with \mirgecom~results (middle) and \software{Ryujin} results (bottom) are shown for magnitude of the density gradient at an advanced simulation time.}\label{fig:forward_step_mirgecom_ryujin}
\end{figure}

The first example is the classic setup of an inviscid flow past a
forward-facing step, with a domain as shown
in~\cref{fig:forward_step_mirgecom_ryujin}. A perfect gas is assumed
($\gamma=1.4$), and adiabatic slip conditions are imposed on the top and bottom
portions of the boundary.  At the inflow, $\Gamma_{\textnormal{in}}$, a Mach $M=3$ inflow speed is set, while no boundary conditions
are imposed at the supersonic outflow $\Gamma_{\textnormal{out}}$.  With velocities non-dimensionalized by speed of sound, the flow is
initialized with velocity $(u,v)=(3,0)$.

The mesh has $\sim$1.6M and $\sim$1M degree-1 TPEs for \mirgecom{} and \software{Ryujin}, respectively (totaling $\sim$6.6M and $\sim$4.1M DOFs for each scalar quantity). A bow shock quickly forms due to the step and grows to reflect from the top boundary, creating a Mach stem and set of reflected shocks.

\Cref{fig:forward_step_mirgecom_ryujin} shows the magnitude of the density gradient,
scaled by the maximum value,
at a simulation time after the Mach stem has
formed. Both \mirgecom{} and \software{Ryujin} show the strong bow shock ahead of the step, as
well as a Mach stem caused by reflection from the top boundary. Shock
reflections from the triple point and top/bottom walls are also visible. Slight
differences can be observed in the vortices shed from the triple point, the
intersection of the bow shock and Mach stem.
There is a close agreement in the resulting flows produced by \mirgecom{} and
\software{Ryujin}, and the differences are attributed to differences in the
underlying discretizations, e.g. continuous Galerkin (CG) in the case of
\software{Ryujin}, a different limiter, and the use of artificial viscosity.
Moreover, \software{Ryujin} solves the compressible Euler equations, while \mirgecom{} solves
the full compressible Navier-Stokes equations.

\begin{figure}[!thp]
\begin{tikzpicture}[x=276.5pt,y=276.5pt]   %
  \begin{scope}[shift={(0pt, 0pt)}]
    \draw[dashed,black!40] (0.5,0) -- (0.5,0.5);
    \draw[black,line width=0.5,line cap=rect] (0,0) -- (0,0.5) -- (1,0.5) -- (1,0) -- cycle;
    \draw[fill=white] (0,0) rectangle (1,0.005);
    \draw[pattern=north east lines, pattern color=black!50] (0,0) rectangle (1,0.005);

    \node[anchor=north] at (0.2,0.5) {adiabatic slip BC};
    \node[anchor=north,rotate=90] at (0,0.3) {adiabatic slip BC};
    \node[anchor=north,rotate=-90] at (1,0.3) {adiabatic slip BC};
    \node[anchor=south] at (0.2,0.0) {adiabatic noslip BC};
    \node at (0.25, 0.25) {$\Omega_{\textnormal{L}}$};
    \node at (0.75, 0.25) {$\Omega_{\textnormal{R}}$};

    \draw[{Stealth[]}-{Stealth[]},black!20,shorten <=0.25pt,shorten >=0.25pt] (0,0.1) -- (1, 0.1) node[pos=0.2,above,black!10] {$1.0$};
    \draw[{Stealth[]}-{Stealth[]},black!20,shorten <=1.0pt,shorten >=0.25pt] (0.8,0) -- (.8, 0.5) node[pos=0.8,left,black!10] {$0.5$};
  \end{scope}
\end{tikzpicture}

      \begin{tikzpicture}
        \node[inner sep=0pt] (image) at (0,0) {
          \includegraphics[width=0.7\textwidth,frame]{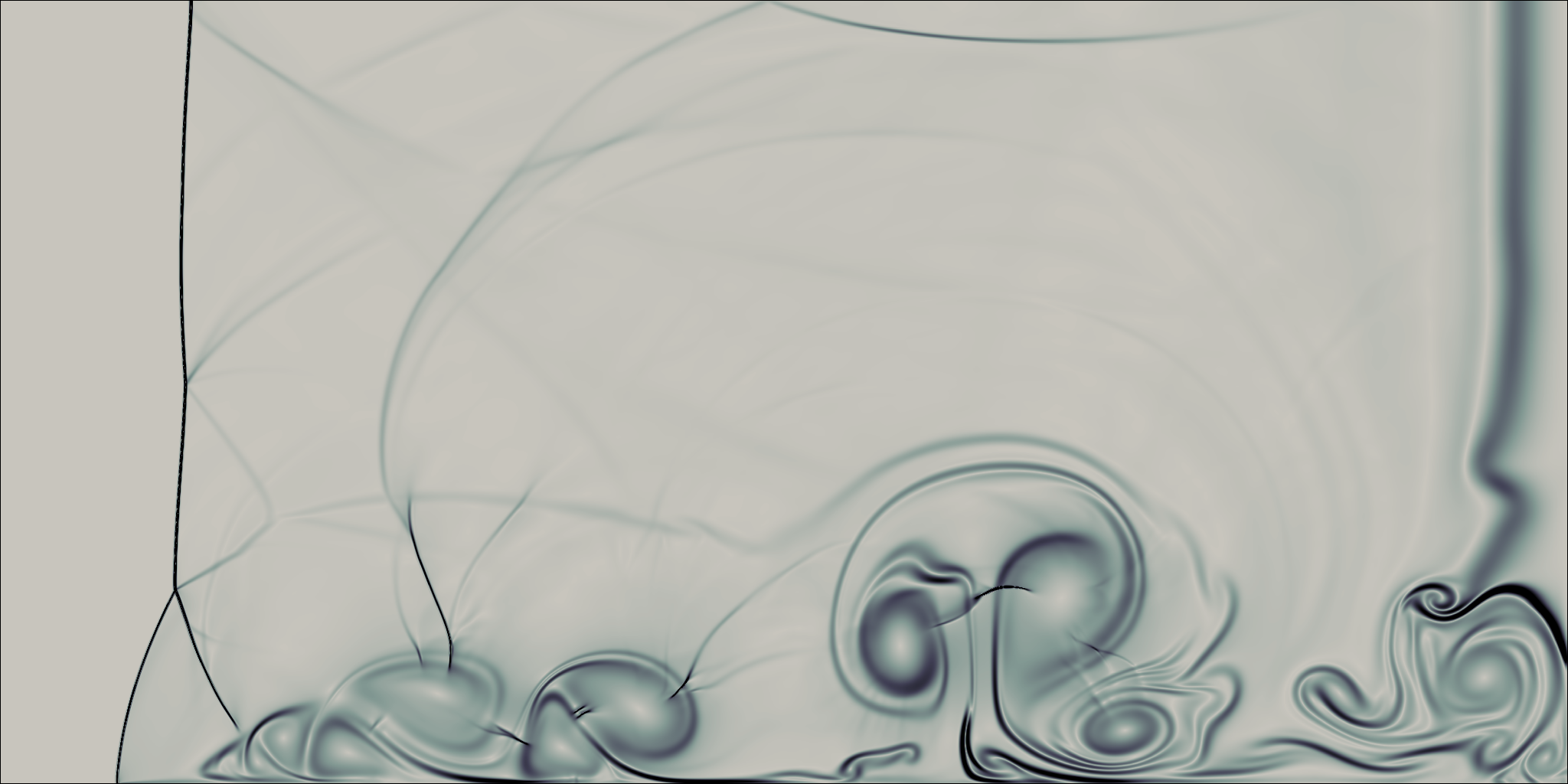}
        };
        \node at (0,2) {\mirgecom{}};
      \end{tikzpicture}
      \begin{tikzpicture}
        \node[inner sep=0pt] (image) at (0,0) {
          \includegraphics[width=0.7\textwidth,frame]{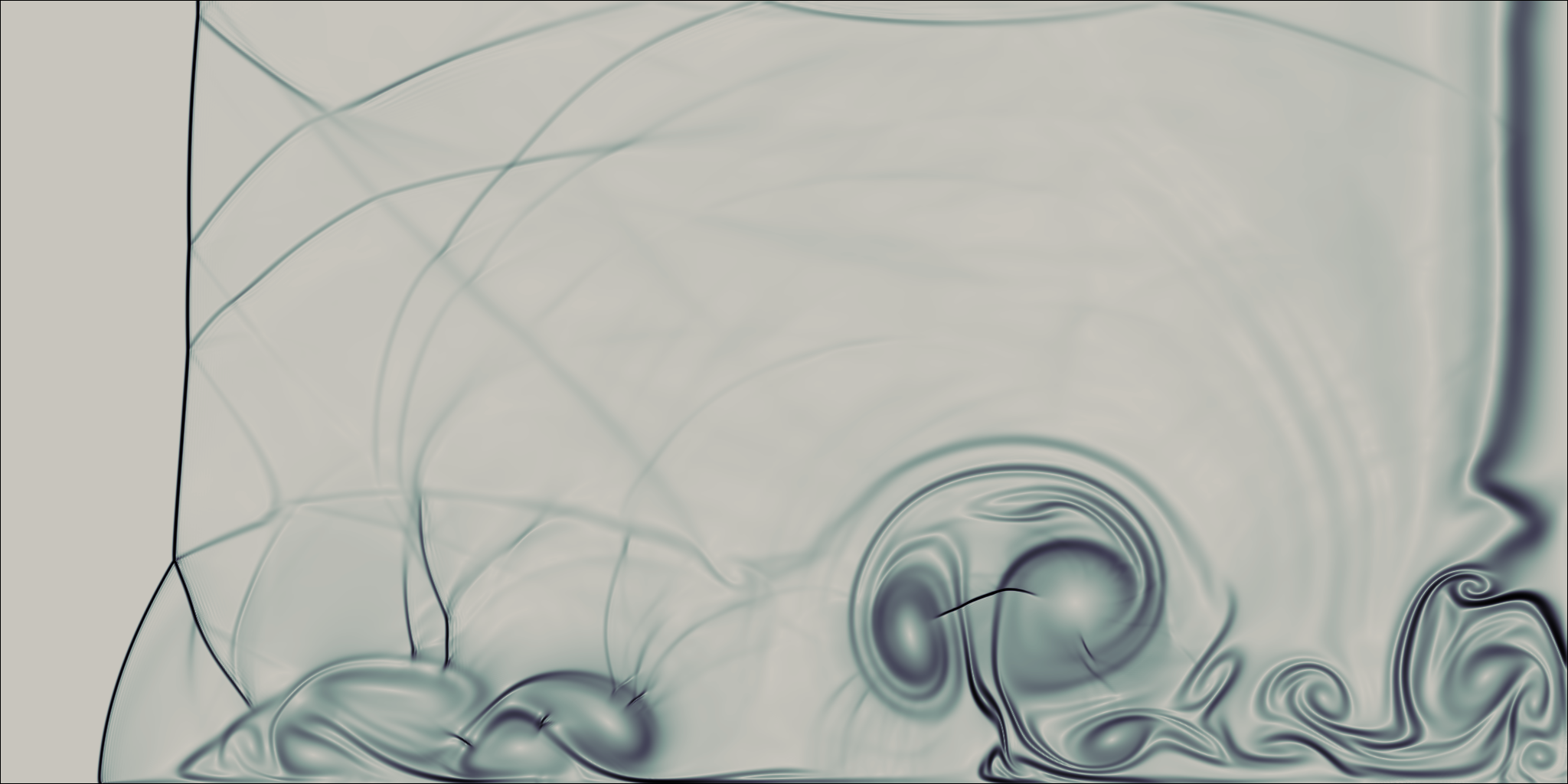}
        };
        \node at (0,2) {\software{Ryujin}};
      \end{tikzpicture}
  \Description{Schematic of the domain for a shock.  The figure shows the dimensions to be 1 by 0.5. Two figures, one for MIRGE-Com and one for Ryujin, show similar profiles of a developed flow, with shocks and small fluid features.}
  \caption{Shock wave interaction with a boundary layer. Problem setup is shown at top with \mirgecom~results (middle) and \software{Ryujin} results (bottom) are shown for magnitude of the density gradient at an advanced simulation time, after shock reflection.}\label{fig:shock_mirgecom_ryujin}
\end{figure}
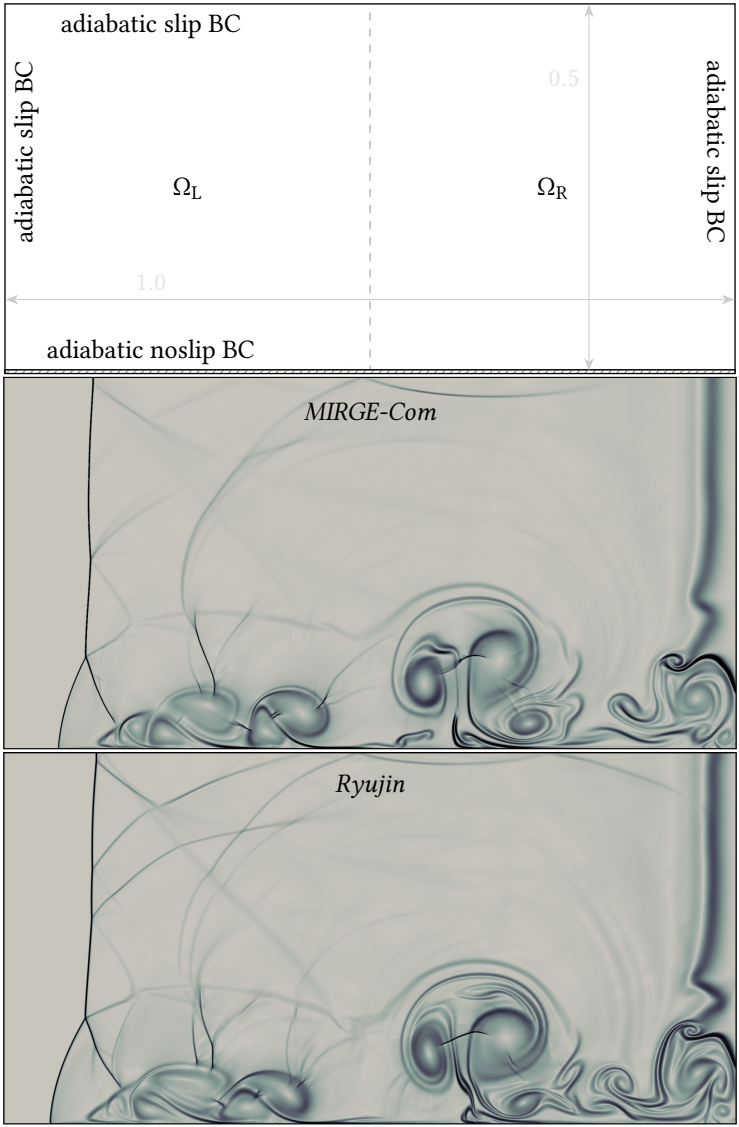

Next, we consider the interaction of a shock wave with a boundary layer. The
domain is described in~\cref{fig:shock_mirgecom_ryujin}, and a perfect gas is
assumed ($\gamma=1.4$). The fluid is initialized at rest, $(u,v)=(0,0)$, with a
discontinuity in the density, with
$\rho_{\textnormal{L}}=120.0$ in $\Omega_{\textnormal{L}}$ and
$\rho_{\textnormal{R}}=1.2$ in $\Omega_{\textnormal{R}}$, and with
pressure prescribed by the isentropic flow relations: $P=\rho/\gamma$.
Adiabatic slip conditions are imposed on the top and sides, while an adiabatic
no-slip condition is imposed on the bottom of the domain. These conditions
generate a shock wave that travels from left to right towards a reflective
wall. A boundary layer forms along the bottom wall as the shock traverses over
it. Upon reflection from the end wall, the shock interacts with the boundary
layer, depositing vorticity due to the misalignment of the density and pressure
gradients.

The \mirgecom{} mesh has 2M TPEs with degree-1 bases (8M DOFs), while the \software{Ryujin} mesh has $\sim$2.1M elements (8.4M DOFs).
\Cref{fig:shock_mirgecom_ryujin} shows the magnitude of the density gradient,
scaled by the maximum value, at a simulation time after the fluid has
developed.

The use of different models and discretizations, makes a direct comparison of
performance difficult.  Nevertheless, the raw numbers are informative and we
report the right-hand-side (RHS) evaluation times for both applications on an Apple
M2: the throughput for \software{Ryujin} is approximately 3.8M DOFs
per second, while the throughput for \mirgecom{} is approximately 1.6M
DOFs per second. \mirgecom{} also benefits substantially from its
portability and device-readiness. Exercising this benchmark on an AMD MI300A
device, for example, \mirgecom{} RHS throughput was 100.4M DOFs per second.

A natural counterpart to performance is productivity, and one beneficial feature
of \mirgecom{} is that extending and modifying the model is relatively straightforward,
due to its array-centric design and the expressiveness of Python.
To quantify productivity, one metric is total lines of code,
for example $\sim$48.4k lines in the case of Ryujin (without \software{deal.ii}) and
$\sim$21.1k in \mirgecom{} (without subpackages such as \software{pytato}).  In
the end, however, lines of code does not fully capture the ease of adding new
features (in Python) and the value of device-ready portability.

\subsection{Complex Geometry Simulation}\label{sec:ceesd-prediction}

\begin{figure}[!thp]
\centering
\includegraphics[height=1.0in]{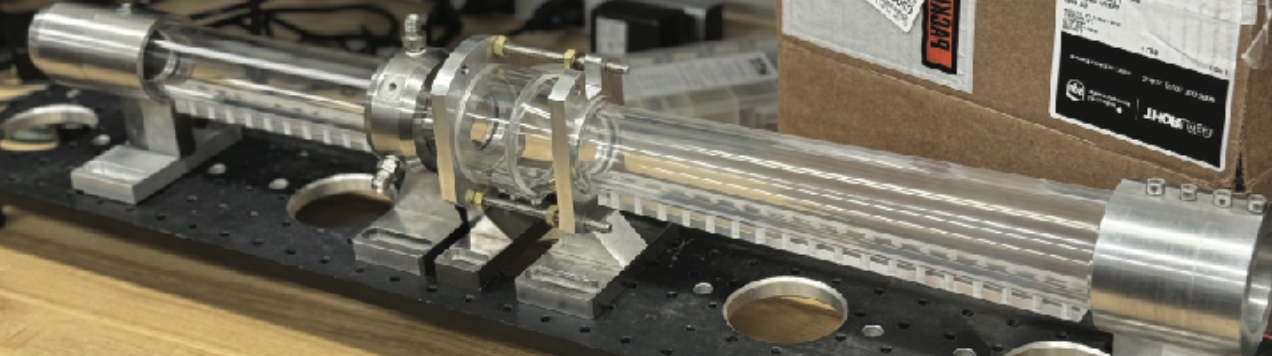}\\
\includegraphics[height=1.5in]{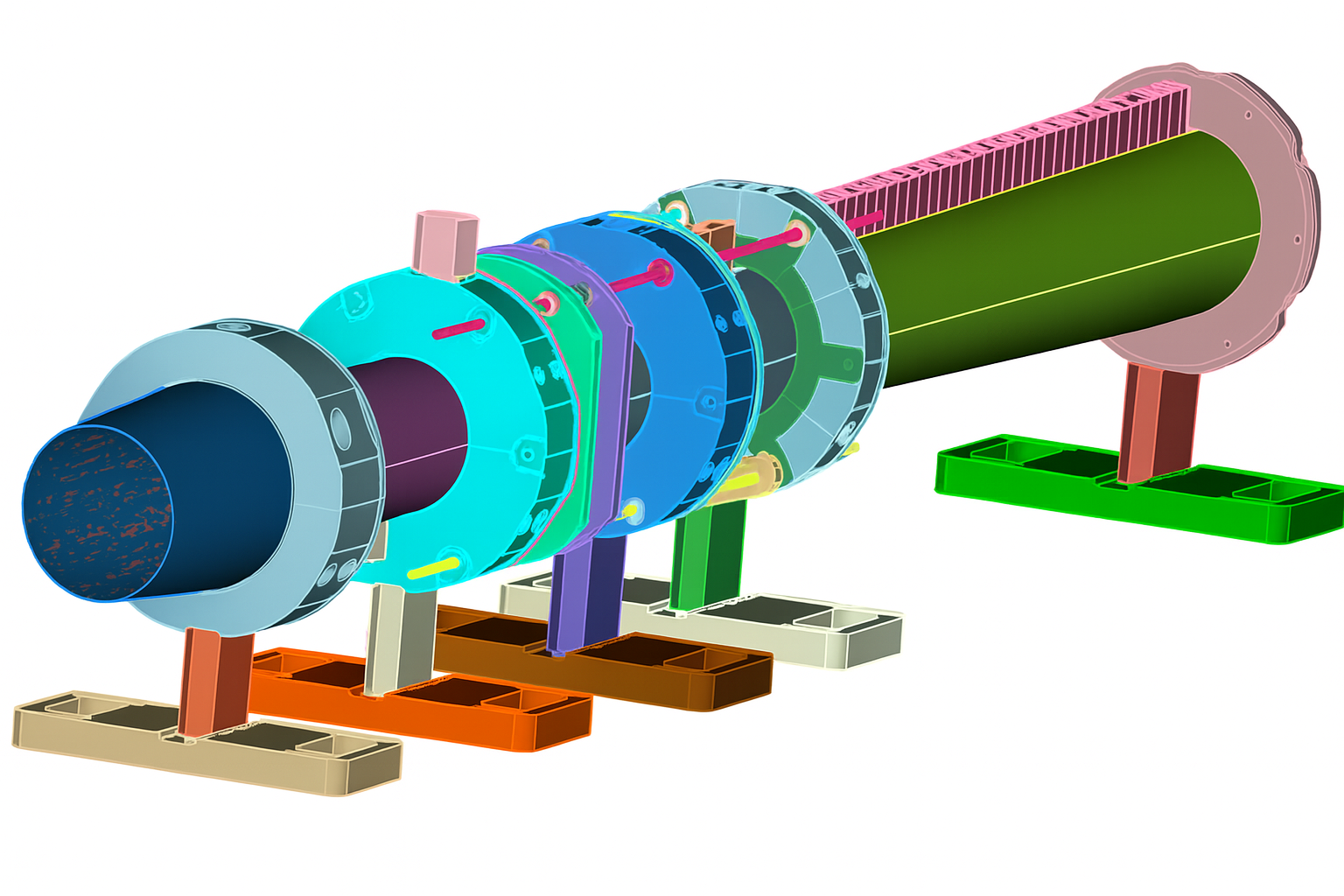}\\
\includegraphics[height=1.5in]{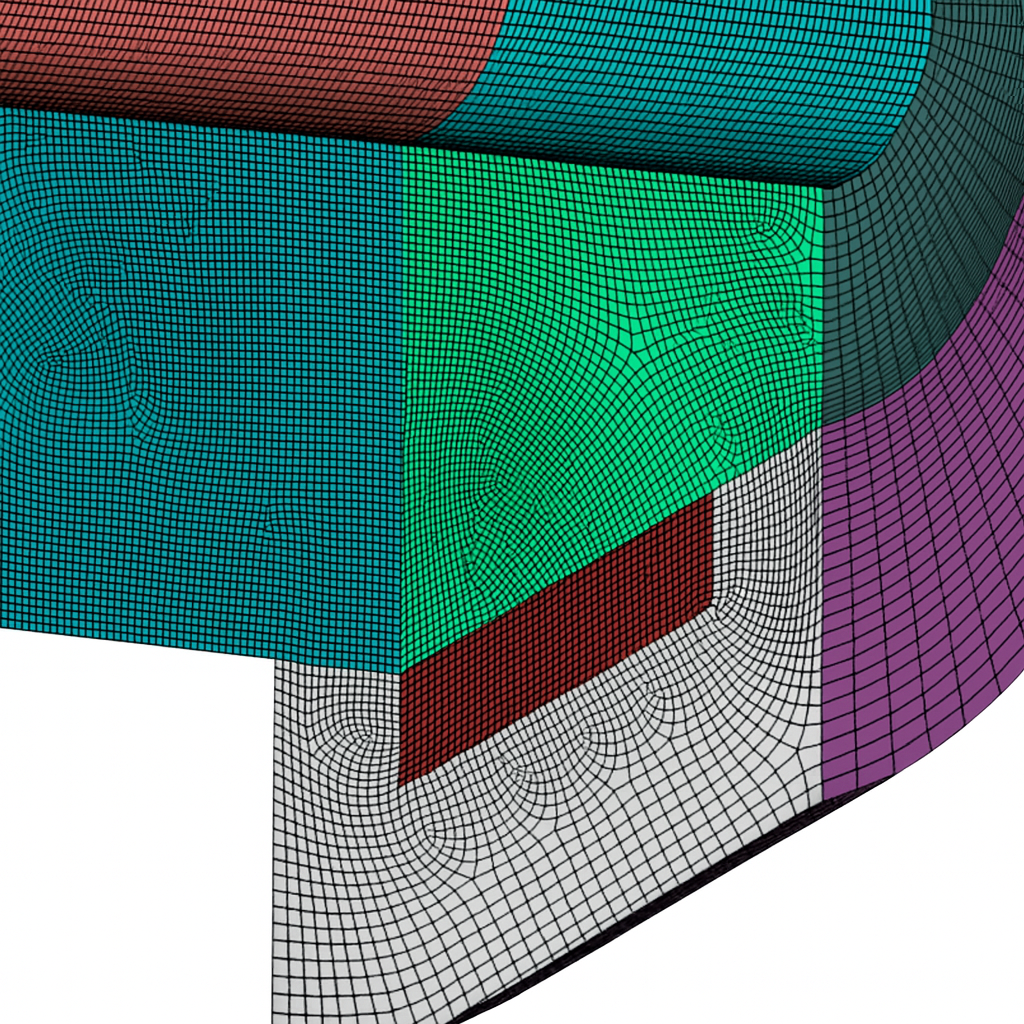}
\Description{A photo showing a long cylindrical experimental setup.  A CAD depiction of the image.  A slice of the hexahedral mesh in 3D.}
\caption{(top) ACTII supersonic combustion apparatus [image credit: Izzi Gessman, Illinois]
         (middle) raw CAD model [CAD Credits: Qili Liu, Izzi Gessman, Mike Anderson, Illinois]; and
         (bottom) a section of the computational mesh produced using Coreform~\cite{coreform}.
        }\label{fig:predprep}
\end{figure}

\mirgecom{} is used by CEESD in a real-world science application to predict the behavior of a lab-scale scramjet combustor experiment conducted in the Arc Heated Combustion Tunnel-II (ACTII) facility~\cite{actii_1, actii_2} at Illinois. The ACTII experimental apparatus and its computational counterpart are shown in~\cref{fig:predprep}.
In this simulation, the mesh of 34M TPEs resulted in over 3B DOFs for a three-dimensional gas-domain simulation.  The computation and mesh are distributed among processing elements using Metis~\cite{metis}.

\begin{figure}[!thp]
  \centering
  \includegraphics[width=0.8\textwidth]{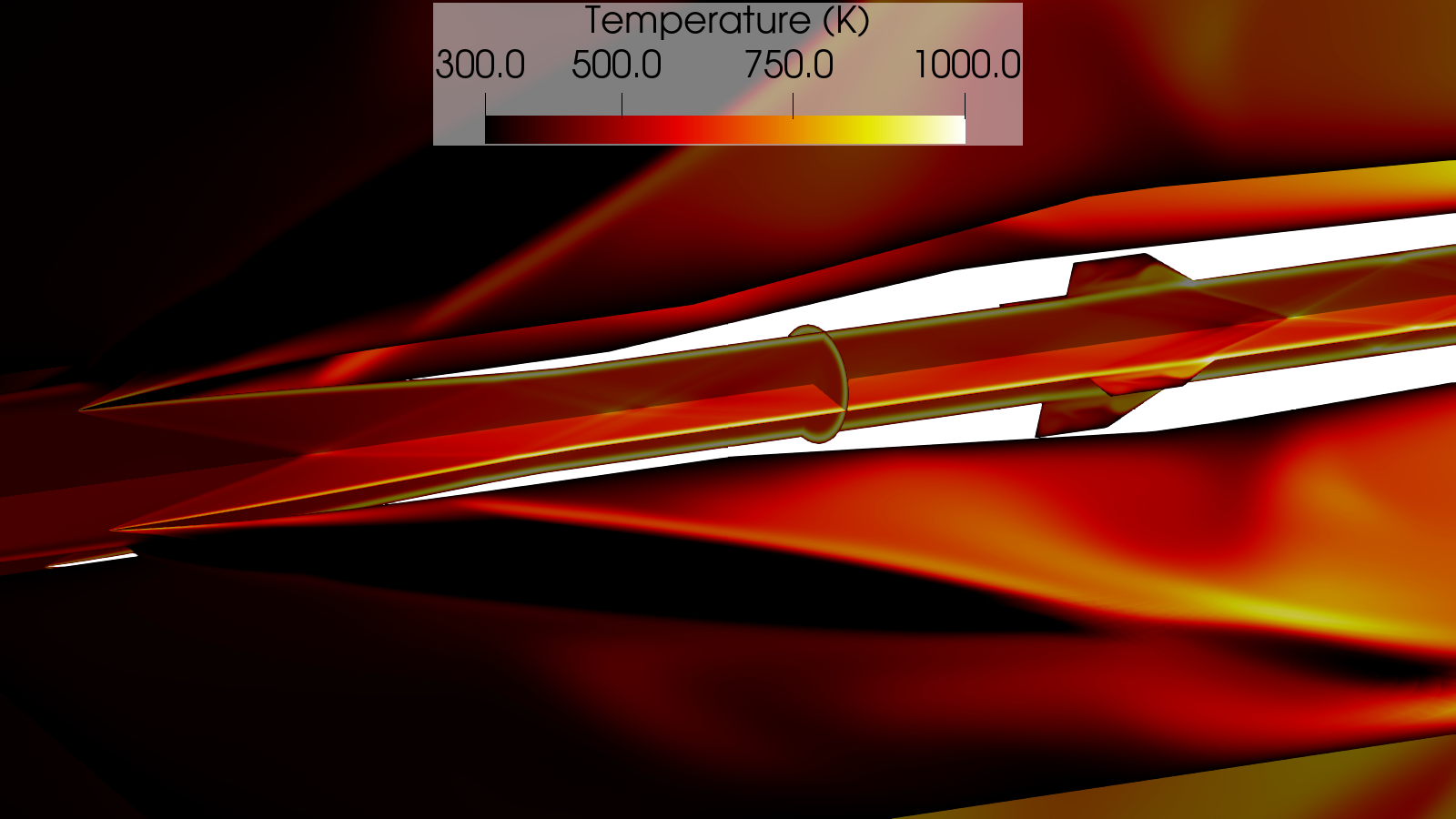}
  \Description{A 3D image of the fluid flow in the domain showing the temperature.}
  \caption{Section of the scramjet combustor simulation using 256 AMD MI300A APUs on the Tuolumne platform at LLNL.}\label{fig:predsim}
\end{figure}

The simulation was conducted using (based on convenience and availability) a variety of computational platforms, while many scoping simulations and capability developments were conducted on laptops (both Intel/Linux and Apple).  The development and scoping runs support large-scale simulations which run on HPC resources at DOE National Laboratories, particularly the LLNL Tuolumne system, which is equipped with AMD MI300A accelerators. A snapshot of the capstone simulation is shown in~\cref{fig:predsim}.

The parallel scalability of the simulation is important as the trend can be
used to estimate both the computational resource required and the expected
time-to-solution (TTS) for a given simulation. Here, we assess the performance
of the large-scale scramjet simulation on the Tuolumne platform at LLNL, measuring
both \textit{weak} and \textit{strong} parallel scalability.

\begin{figure}[!tb]
  \centering
  \includegraphics{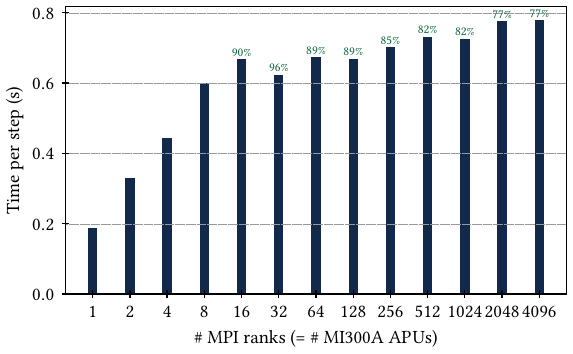}
  \Description{A bar chart with the time per time step showing slight growth as more ranks are used.}
  \caption{Weak parallel scaling on Tuolumne (AMD MI300A) for the scramjet baseline simulation.
  The weak scaling parallel efficiency for $n$ ranks versus $8$ ranks ($100\cdot t_8/t_n$) is given as a percentage.}\label{fig:scaling_weak_1dpart_metis}
\end{figure}

\textbf{Weak scaling} measures the performance of a larger problem given a larger computational resource. 
\Cref{fig:scaling_weak_1dpart_metis} shows weak scaling results for a 1D gas shock impacting a gas-material interface in a three-dimensional domain, which offers a baseline for the full scramjet simulation. The weak scaling results demonstrate that \mirgecom{} can achieve near-ideal scaling to full machine scale for this leadership-class HPC platform.

\begin{figure}[!tb]
  \centering
  \includegraphics{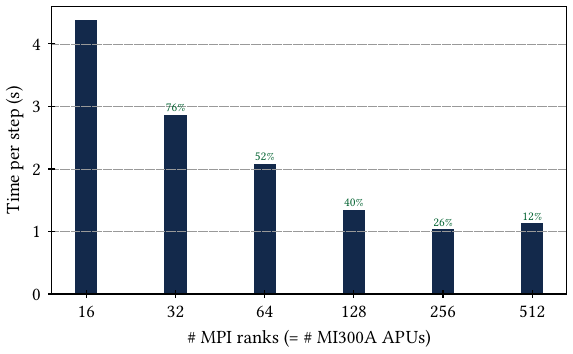}
  \Description{A bar chart with the time per time step showing a decrease as more ranks are used until 256.}
  \caption{Parallel strong scaling on Tuolumne (AMD MI300A) for the scramjet simulation.
           The strong scaling parallel efficiency for $n$ ranks versus $16$ ranks ($100\cdot 16 \cdot t_{16}/(n \cdot t_{n})$) is given as a percentage.}\label{fig:scaling_strong_metis}
\end{figure}

\textbf{Strong scaling} measures the performance of fixed-size problem faster
given a larger computational resource.  For this test we fix the mesh size in
the simulation at $\sim$3M elements (or $\sim$1.26B DOFs),
resulting in $\sim$80M DOFs per rank to $\sim$2.5M DOFs per rank as we scale
the number of GPUs from 16 to 512. \Cref{fig:scaling_strong_metis} shows strong
scaling, where we observe good scaling out to 256 ranks for this run, which was
more than sufficient for conducting the desired simulations with reasonable
turn-around time.

The \mirge{} process often obscures the code-to-kernel connection, making it a challenge
to correlate the resulting kernels with expected performance of specific DG operations.
The \mirge{} infrastructure transforms the application code dynamically
at runtime, and the transformations are problem and configuration specific. Our expectation
for low-order ($p \le 2$) DG is that the work is dominated by vector (BLAS-1) and
matrix-vector (BLAS-2) \cite{blas} operations, yielding kernels with low arithmetic intensities
and a memory bound execution.

\begin{figure}[htbp] \centering
  \includegraphics{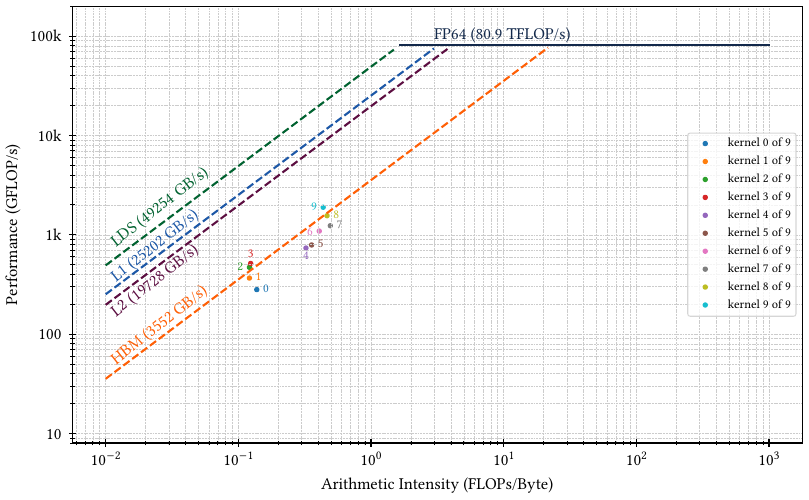}
  \Description{A log-log scale plot showing sloped line representing the HBM.  The 10 kernels are represented as near the HBM line.}
  \caption{Empirical roofline on Tuolumne (AMD MI300A) for top 10 most
  expensive \mirgecom{} kernels in the scramjet simulation running $p=1$ simplices.}\label{fig:prediction-roofline}
\end{figure}

To evaluate the absolute performance of \mirgecom{}, we examine the roofline
performance~\cite{roofline-original} for the most expensive kernels (i.e. those in which the
application spends the most time) generated by the \mirge{} process for the scramjet simulation
running $p=1$ simplicial elements.  \Cref{fig:prediction-roofline} presents an empirical
roofline model constructed from the ten most expensive kernels, using the AMD roofline
tool~\software{rocprof-compute}; the tool empirically evaluates the
arithmetic intensity and measures the running bandwidth for the
device at runtime. The roofline plot illustrates that the
kernels generated for this problem run close to the expected bandwidth for main high-bandwidth
memory (HBM).  Kernels exceeding the HBM performance have a higher amount of
data reuse (e.g., cache-coherence), while those falling significantly short of
HBM bandwidth indicate kernels that could potentially be improved.

\section{Conclusion}

The \mirge{} approach for array-based
high-performance computing builds on the concept of \emph{lazy evaluation}
of array-based computation. By representing computation as array dataflow graphs (ADFGs),
it enables transformations and optimizations that are impractical in a
purely eager execution model. These graphs are lowered
to a scalar IR, where traditional compiler-style transformations such as loop
fusion become possible. From there, target-specific OpenCL kernels are
generated and executed. The framework also integrates distributed execution
directly into the ADFG through specialized communication nodes, allowing
large-scale computations to be expressed in a unified way.  Demonstrations of simulations show it provides platform portability and performance loosely comparable to other finite-element flow simulation tools.  

\begin{acks}
  This material is based in part upon work supported by the \grantsponsor{doe}{Department of Energy, National Nuclear Security Administration}, under Award Number \grantnum{doepsaap}{DE-NA0003963}
  as well as by the \grantsponsor{nsf}{National Science Foundation} under Award Number
  \grantnum{nsfloopy}{1931577}.
\end{acks}

\printbibliography
\end{document}